\documentclass[10pt,twocolumn,twoside]{IEEEtran}
\usepackage{cite}
\usepackage{amsmath,amssymb,amsfonts}
\usepackage{algorithmic}
\usepackage{graphicx}
\usepackage{textcomp}
\usepackage{amsthm}

\usepackage{todonotes}

\newtheorem{assum}{Assumption}
\newtheorem{definition}{Definition}
\newtheorem{lemma}{Lemma}
\newtheorem{thm}{Theorem}
\newtheorem{prop}{Proposition}
\newtheorem{cor}{Corollary}

\newcommand{\profit}{\phi}

\usepackage{comment}
\specialcomment{proofsketchcomment}{}{}

\specialcomment{proofcomment}{}{}

\includecomment{proofcomment}
\excludecomment{proofsketchcomment}

\newcommand{\change}[1]{{\color{black} #1}}

\newenvironment{proofsketch}{%
  \proof}{\endproof}

\def\BibTeX{{\rm B\kern-.05em{\sc i\kern-.025em b}\kern-.08em
    T\kern-.1667em\lower.7ex\hbox{E}\kern-.125emX}}

\begin{document}
\title{Mixed Autonomy in Ride-Sharing Networks}
\author{Qinshuang Wei, \IEEEmembership{Student Member, IEEE}, Ramtin Pedarsani, \IEEEmembership{Member, IEEE}, and Samuel Coogan, \IEEEmembership{Member, IEEE}

\thanks{Q. Wei and S. Coogan are with the School of Electrical and Computer Engineering, Georgia Institute of Technology, Atlanta, GA 30332 USA (e-mail: {qinshuang,  sam.coogan}@gatech.edu). S. Coogan is also with the School of Civil and Environmental Engineering, Georgia Institute of Technology.  This work was funded in part by the National Science Foundation under grant 1749357.}
\thanks{ R. Pedarsani is with the Department of Electrical Engineering, University of California, Santa Barbara, CA 93106 USA (e-mail: ramtin@ece.ucsb.edu).}}

\maketitle

\begin{abstract}
We consider ride-sharing networks served by human-driven vehicles (HVs) and autonomous vehicles (AVs). We propose a model for ride-sharing in this mixed autonomy setting for a multi-location network in which a ride-sharing platform sets prices for riders, compensations for drivers of HVs, and operates AVs for a fixed price with the goal of maximizing profits. When there are more vehicles than riders at a location, we consider three vehicle-to-rider assignment possibilities: rides are assigned to HVs first; 
rides are assigned to AVs first;  
rides are assigned in proportion to the number of available HVs and AVs. Next, for each of these priority possibilities, we establish a nonconvex optimization problem characterizing the optimal profits for a network operating at a steady-state equilibrium. We then provide a convex problem which we show to have the same optimal profits, allowing for efficient computation of equilibria, and we show that all three priority possibilities result in the same maximum profits for the platform. Next, we show that, in some cases, there is a regime for which the platform will choose to mix HVs and AVs  in order to maximize its profit, while in other cases, the platform will use only HVs or only AVs, depending on the relative cost of AVs. For a specific class of networks, we fully characterize these thresholds analytically and demonstrate our results on an example.

\end{abstract}

\section{Introduction}
\label{sec:introduction}

Ride-sharing platforms, also known as transportation network companies, have become commonplace due factors such as high costs of car ownership, lack of parking, and persistent traffic congestion \cite{Mitchell2010, Board2016, Hass-Klau2007, Javid2016, Mcbain2018}. Traditionally, rides are provided by drivers who use their personal vehicle to provide service.  However, ride-sharing platforms are likely to incorporate autonomous vehicles (AVs) into their fleets in the near future \cite{Litman2017}.   
Nonetheless, significant technological and regulatory hurdles remain before ride-sharing platforms could transition to 100\% autonomous fleets \cite{fagnant2015preparing, guerra2016planning}. 
Therefore, it is likely that ride-sharing platforms will initially adopt a \emph{mixed} framework in which AVs operate alongside conventional, human-driven vehicles (HVs) \cite{Boesch2016, grush2018end, Conger:2018ad}.

Existing research in ride-sharing has largely focused on two ends of the autonomy spectrum. On one end are futuristic \emph{mobility-on-demand} systems consisting of only AVs \cite{Zhang2015, Rigole2014, Zhang2015a, Zhang2016, fagnant2018dynamic}. 
On the other end, models of rider and driver behavior in conventional ride-sharing markets with only HVs and no AVs have been considered in \cite {Bimpikis2018, Banerjee2015, Cachon2017, Banerjee2016}. 

In this paper, we study the transition from traditional ride-sharing networks to totally automated mobility-on-demand systems. In particular, we extend the  model proposed in \cite{Bimpikis2018}, which did not consider AVs, to the mixed autonomy setting under several assumptions on the vehicle-to-rider assignment possibilities, and we analyze the resulting models. We consider a network consisting of multiple locations, and potential riders arrive at these locations with desired destinations. The ride-sharing platform sets prices for riders and compensation to drivers of HVs.
 In addition, the platform has the option to deploy AVs for a fixed cost. Introducing AVs leads to an important assignment choice that must be made: if both an AV and an HV are available to serve a rider, which receives preference? We consider three possible assignment rules: AVs always receive priority (AV priority); HVs always receive priority (HV priority); and priority is determined in proportion to the number of available AVs and HVs at each location (weighted priority).

We focus on the equilibrium conditions that arise in the resulting mixed autonomy deployment when the platform seeks to maximize profits.
We summarize our main findings as follows: 
1) In all three priority assignments, the equilibrium conditions lead to a non-convex optimization problem. Nonetheless, we develop an alternative convex problem from which an optimal solution to the original non-convex problem can be recovered. 
2) We find that, surprisingly, all three priority schemes result in the same maximum profits for the platform. This is because, at an optimal equilibrium, we show that all vehicles are assigned a ride and thus the  priority assignment choice is immaterial at the optimal equilibrium. 
3)  Lastly, we consider the ratio of AVs to HVs that will be deployed by the platform in order to maximize profits for various operating costs of AVs. 
We show that, in some cases, there is a regime for which the platform will choose to mix HVs and AVs vehicles in order to maximize profits, while in other cases, the platform will use only HVs or only AVs, depending on the relative cost of AVs. For a specific family of networks, we fully characterize these thresholds analytically. 

\change{The main contributions of this paper are therefore two-fold. First, we develop a new model for studying the emergence of AVs in ride-sharing networks. This model contributes substantial modifications to the foundational model developed in \cite{Bimpikis2018} in order to allow for the presence of AVs. Second, we conduct a detailed theoretical study of the resulting model focusing on the optimal profits obtainable by a ride-sharing platform that deploys AVs.} To the best of our knowledge, the present paper is the first to provide a formal framework for understanding and quantifying the impact of integrating AVs into ride-sharing fleets\footnote{This paper extends our preliminary work \cite{wei2019routing}, which only considered AV priority assignment, and the theoretical results in \cite{wei2019routing} are limited to a specific class of networks.}.

The remainder of this paper is organized as follows. Section \ref{sec:model} provides the model definitions, and Section \ref{sec:optimization-problem} poses the problems of profit maximization as non-convex optimization problems. Section \ref{sec:alt-orig-optim} proposes an alternative convex optimization problem that provides the same optimal profits 
and from which a solution to the original problem can be recovered. In Section \ref{sec:D-A-model-relation}, we study the relation between the AV and HV priority assignments and show that they achieve the same optimal profits. Due to its asymmetry to the AV and HV priority assignments, weighted priority assignment is introduced and studied separately in Section \ref{sec:weighted-model}. 
Section \ref{sec:numerical-study} studies a particular class of networks and fully quantifies the profit maximizing equilibrium conditions.  Concluding remarks are provided in Section \ref{sec:conclusions}\footnote{Complete proofs are contained in the extended version arXiv:1908.11711, available at http://arxiv.org/abs/1908.11711.}. 
\begin{proofsketchcomment}
\end{proofsketchcomment}

\section{Problem Formulation}
\label{sec:model}
We consider an infinite horizon discrete time model of a ride-sharing network that extends the model recently proposed in \cite{Bimpikis2018} to accommodate a mixed autonomy setting with autonomous vehicles (AVs) and human-driven vehicles (HVs). 
The network operator or \emph{platform} determines prices for rides and compensations to drivers within the network. 
The price of a ride may differ among locations, but does not depend on the desired destination of each rider.

\change{In this paper, we focus on equilibrium conditions that arise when the demand pattern of riders is stationary.  
For example, for several hours in the early evening on weekends, there might be steady and predictable demand for rides from residential areas to entertainment districts. 
An alternative direction of research is to consider, for example, the transient effects of changing demand over time. While the model developed below could be utilized in such a context, we focus only on stationary demand and the resulting equilibrium conditions here.}

With these considerations in mind, we study the potential benefits of adding AVs to the network to maximize the profit potential for the platform.

\subsection{Model Definition}

We now formalize the mixed autonomous ride-sharing network described above. 

\underline{\emph{Riders.}} Among a network of $n$ equidistant locations, a mass of $\theta_i$ potential riders arrives at location $i\in\{1,2,\ldots,n\}$ in each period of time. Throughout, when indices are omitted from a summation expression, it is assumed the summation is over all locations $1$ to $n$. A fraction $\alpha_{ij}\in[0,1]$ of riders at location $i$ are traveling to location $j$ so that $\sum_{j} \alpha_{ij} = 1$ for all $i$. We assume $\alpha_{ii}=0$ for all $i$ and construct the $n$-by-$n$ adjacency matrix $\mathbf{A}$ as $[\mathbf{A}]_{ij} = \alpha_{ij}$ where $[\mathbf{A}]_{ij} $ denotes the $ij$-th entry of $\mathbf{A}$.

\underline{Human-driven vehicles (\emph{HVs}).} After each time period, a driver exits the platform with probability $(1-\beta)$ \change{and serves another ride with probability $\beta$} where $\beta \in(0,1)$. 
Thus, a driver's expected lifetime in the network is $(1-\beta)^{-1}$. Each driver has an outside option of earning $\omega$ over the same lifetime. 

\underline{Autonomous vehicles (\emph{AVs}).} The platform can choose to operate an AV in the network for a fixed cost of $s$ each time-step. Thus, $k= s(1-\beta)^{-1}/\omega$ is the ratio of the cost of operating an AV for the equivalent time of a driver's expected lifetime to the outside option earnings. Unlike HVs, it is assumed that AVs are in continual use and do not leave the platform. 

\underline{\emph{Platform.}} The platform sets a price $p_i$ for a ride from location $i$ and correspondingly compensates a driver with $c_i$ for providing a ride at location $i$. The continuous cumulative distribution of the riders' willingness to pay is denoted by $F(\cdot)$ with support $[0,\bar{p}]$. That is, when confronted with a price $p$ for a ride, a fraction $1-F(p)$ of riders will accept this price, and the remaining $F(p)$ fraction will balk and leave the network without requesting a ride. Note that $\theta_i (1-F(p_i ))$ is then the effective demand for rides at location $i$.

\change{The description of the riders, HVs, and the platform is the same as that presented in \cite{Bimpikis2018}. In this work, we also introduce AVs as described above. As developed below, this addition substantially alters how the model behaves and is analyzed as compared to \cite{Bimpikis2018}.} In addition, we make the following assumption throughout.

\begin{assum}
\label{assum:all}
The network's demand pattern is stationary, \emph{i.e.}, $\mathbf{A}$ and $\theta_i$ are fixed for all $i$. Moreover, the directed graph defined by adjacency matrix $\mathbf{A}$ is strongly connected and $\theta_i > 0$ for all $i\in\{1,\ldots, n\}$, $n\geq 2$.
\end{assum}

In summary, the system consists of a \emph{platform} that sets prices, \emph{riders} that request rides among locations,  \emph{HVs} that seek to maximize their compensation, and \emph{AVs} managed by the platform alongside the drivers.

\subsection{HV and AV Priority Assignments}
The number of riders willing to pay the platform's price may be less than, equal to, or greater than the total number of HVs and AVs available at that location. When it is greater than the total number of vehicles, some riders will not be served and will leave the network. When it is less than the total number of vehicles, the platform must decide how to assign riders to vehicles. \change{Resolving this priority assignment problem is one of the main challenges presented by the model defined above as compared to the model with no AVs as proposed in \cite{Bimpikis2018}. When no AVs are present, it is assumed that riders are arbitrarily assigned to drivers and any remaining HVs choose to reroute to the location of highest expected earnings. In contrast, }in this paper, we consider several priority assignments.

The first priority assignment, called \emph{HV priority}, assigns riders to HVs before AVs and is appropriate if, \emph{e.g.}, the platform views HVs as customers that should be accommodated and given preference over AVs. We also consider an \emph{AV priority} assignment in which AVs are assigned rides before HVs. This priority assignment is appropriate if, \emph{e.g.}, the platform views HVs only as a supplement when insufficient AVs are available. 
In Section \ref{sec:weighted-model}, we consider a third, intermediate \emph{weighted} priority assignment that assigns rides in proportion to the availability of vehicles, but we defer its definition and analysis until later.

We sometimes refer to the above defined model under any of the three priority assignments as a \emph{mixed autonomy deployment}. For comparison, the \emph{HV-only deployment} is obtained by assuming no AVs at any location. An \emph{HV-only deployment} may arise by the choice of a profit-maximizing platform if the platform decides not to use any AVs; alternatively, we may consider an HV-only deployment by enforcing the constraint of no AVs at any locations, in which case it is referred to as a \emph{forced} HV-only deployment. Similarly, the \emph{AV-only deployment} is obtained from the mixed autonomy deployment when there are no HVs at any locations, and a \emph{forced} AV-only deployment arises when this condition is enforced as a constraint on the system.

\subsection{Equilibrium Definition for HV Priority Assignment}

We now turn to the equilibrium conditions of the above model that are induced by the stationary demand as characterized in Assumption \ref{assum:all} and by fixed prices and compensations set by the platform. An equilibrium for the system is a time-invariant distribution of the mass of riders, HVs, and AVs at each location satisfying certain equilibrium constraints, as formalized next; all variables are understood to refer to an equilibrium and therefore no time index is included.

We consider first HV priority assignment. Let $x_i$ denote the mass of HVs at location $i$. Recall $\theta_i (1-F(p_i))$ the mass of riders willing to pay for a ride at location $i$. If there are fewer riders than HVs at a location, drivers can relocate to another location to provide service in the next time period. For each $i,j\in\{1,\ldots, n\}$, let $y_{ij}$ denote such drivers at location $i$ who relocate to location $j$ without providing a ride. It follows that
\begin{equation}
  \label{eq:1}
\sum_{j=1}^n y_{ij}= \max\left\{ x_i - \theta_i (1-F(p_i)), 0  \right\}  .
\end{equation}
Further, let $\delta_i$ denote the mass of new drivers who choose to enter the platform and provide service at location $i$ at each time step. At equilibrium, it must hold that
\begin{equation}
  \label{eq:2}
x_i=\beta \left[\sum_{j=1}^n \alpha_{ji} \min \left\{ x_j , \theta_j(1-F(p_j)) \right\} + \sum_{j=1}^n y_{ji} \right] + \delta_i.
\end{equation}
In \eqref{eq:2}, observe that $\min \left\{ x_j , \theta_j(1-F(p_j)) \right\}$ is the total demand the platform serves with HVs at location $j$, and therefore  $\sum_{j} \alpha_{ji} \min \left\{ x_j , \theta_j(1-F(p_j)) \right\}$ is the mass of HVs that find themselves located at $i$ after completing a ride. 

When the demand $\theta_i(1-F(p_i))$ at location $i$ exceeds the mass of available HVs $x_i$, the platform can choose to use AVs to meet this extra demand. Let $z_i$ denote the mass of AVs at location $i$, and for each $i,j\in\{1,\ldots,n\}$, let $r_{ij}$ denote the AVs which do not get a ride at $i$ and are relocated to location $j$. 
Then
\begin{equation}
\label{eq:3}   z_i=\sum_{j=1}^n \alpha_{ji} \min \left\{ z_j , \max \left\{\theta_j(1-F(p_j))-x_j , 0\right\} \right\}+ \sum_{j=1}^n r_{ji}.
\end{equation}
In \eqref{eq:3}, observe that $\min \left\{ z_j , \max \left\{\theta_j(1-F(p_j))-x_j , 0\right\} \right\}$ is the total demand that the platform serves with AVs at location $j$. 
Moreover, $\sum_{j} r_{ji}$  is the mass of AVs which do not get a ride to any other location and are relocated to location $i$. It follows that
\begin{equation}
\label{eq:4}
\sum_{j=1}^n r_{ij} = \max \left\{ z_i - \max \left\{ \theta_i (1-F(p_i))-x_i, 0  \right\}, 0 \right\} .
\end{equation}
Note that\change{, under HV priority assignment,} $\sum_{j} r_{ij}$ depends on $x_i$.

\change{For each location $i$, define the \emph{expected earnings} $V_i$ to be the average total compensation earned by a driver arriving at location $i$. Recall that, for each ride served at location $i$, drivers are compensated $c_i$ and travel to a new location according to the demand pattern $\mathbf{A}$. If a driver does not serve a ride due to insufficient demand, the driver earns no compensation but is free to reroute to the location with highest expected earnings. It thus follows that the expected earnings satisfy the relationship} 
\begin{align}
\nonumber V_i &= \min \left\{\frac{\theta_i (1-F(p_i))}{x_i}, 1 \right\}\left(c_i + \sum_{k=1}^n\alpha_{ik}\beta V_{k}\right) \\
 \label{eq:5} &+ \left(1-\min \left\{\frac{\theta_i (1-F(p_i))}{x_i}, 1 \right\}\right)\beta\max_j V_j
\end{align}
\change{for all locations $i$ where we observe ${\theta_i (1-F(p_i))}/{x_i}$ is the fraction of drivers at location $i$ that serve rides, provided $\theta_i (1-F(p_i))\leq x_i$.}

\change{Since drivers have an outside earnings option of $\omega$, they will enter the network at location $i$ if and only if $V_i\geq \omega$. Moreover, the platform is able to independently adjust each compensation $c_i$, so a profit maximizing platform seeking to minimize  $V_i$ is able to achieve $V_i=\omega$ for all $i$, leading to the following definition.}

\begin{definition}
For some prices and compensations $\{p_i,c_i\}_{i=1}^n$, the collection $\left\{\delta_i,x_i,y_{ij},z_i,r_{ij} \right\}_{i,j=1}^n$ is \emph{an equilibrium under $\{p_i,c_i\}_{i=1}^n$ for HV priority assignment} if \eqref{eq:1}--\eqref{eq:4} is satisfied and $V_i$ as defined in \eqref{eq:5} satisfies $V_i=\omega$ for all $i= 1,\ldots, n$ \change{such that $\delta_i+\sum_{j=1}^n y_{ji}>0$.}
\end{definition}

\subsection{Equilibrium Definition for AV Priority Assignment}

In this subsection, we parallel the development of the previous subsection for AV priority assignment. 
The analogous equilibrium conditions are
\begin{align}
\nonumber x_i &=\beta \left[\sum_{j} \alpha_{ji} \min \left\{ x_j , \max \left\{ \theta_j(1-F(p_j))-z_j,0\right\}  \right\} \right. \\
\label{eq:6} &\quad  \left. + \sum_{j} y_{ji} \right] + \delta_i\\
\label{eq:7}
\sum_{j=1}^n y_{ij}&= \max\left\{ x_i - \max \left\{ \theta_i(1-F(p_i))-z_i,0\right\} , 0  \right\}\\
\label{eq:8}
z_i&=\sum_{j=1}^n \alpha_{ji} \min \left\{ z_j , \theta_j(1-F(p_j)) \right\} + \sum_{j} r_{ji}\\
\label{eq:9}
\sum_{j=1}^n r_{ij} &= \max \left\{0,z_i-\theta_i(1-F(p_i))  \right\}.
\end{align}
In comparing \eqref{eq:6}--\eqref{eq:9} to \eqref{eq:1}--\eqref{eq:4}, notice that AV priority assignment leads to $\sum_j y_{ij}$ dependent on $z_i$ in \eqref{eq:7} whereas $\sum_{j=1}^n r_{ij}$ does not depend on $x_i$ in \eqref{eq:9}.

The expected earning $V_i$ for a driver at location $i$ now has the form
\begin{align}
\nonumber V_i &= \min \left\{\frac{M_i}{x_i}, 1 \right\}\left(c_i + \sum_{k=1}^n\alpha_{ik}\beta V_{k}\right) \\
 \label{eq:5-2} 
 &\qquad + \left(1-\min \left\{\frac{M_i}{x_i}, 1 \right\}\right)\beta\max_j V_j,\\
 \label{eq:5-3}M_i&=\max \left\{ \theta_i(1-F(p_i))-z_i,0\right\}.
\end{align}
Again, the platform chooses compensation such that $V_i=\omega$.

\begin{definition}
For some prices and compensations $\{p_i,c_i\}_{i=1}^n$, the collection $\left\{\delta_i,x_i,y_{ij},z_i,r_{ij} \right\}_{i,j=1}^n$ is \emph{an equilibrium under $\{p_i,c_i\}_{i=1}^n$ for AV priority assignment} if \eqref{eq:6}--\eqref{eq:9} is satisfied and $V_i$ as defined in \eqref{eq:5-2}--\eqref{eq:5-3} satisfies $V_i=\omega$ for all $i= 1,\ldots, n$ \change{such that $\delta_i+\sum_{j=1}^n y_{ji}>0$.}
\end{definition}

\section{Profit-Maximization for HV and AV Priority Assignment}
\label{sec:optimization-problem}
We now consider the problem of maximizing profits at equilibrium. \change{We focus on the equilibrium under prices and compensations $\{p_i,c_i\}_{i=1}^n$. This analysis is reasonable when there are large populations of HVs, AVs and riders during periods of stationary rider demand. In this case, the equilibrium captures the flow constraints in \eqref{eq:1}--\eqref{eq:4} or \eqref{eq:6}--\eqref{eq:9} and the drivers' earnings constraints in \eqref{eq:5} or \eqref{eq:5-2}--\eqref{eq:5-3}.} We first consider profit maximization with HV priority assignment and then with AV priority assignment.
Under HV priority assignment, maximizing the aggregate profit across the $n$ locations subject to the system’s equilibrium constraints yields the following optimization problem:
\begin{align}
\nonumber \max_{\{p_i,c_i\}_{i=1}^n}&\sum_{i=1}^n \left[ \min \left\{x_i+z_i, \theta_i (1-F(p_i)) \right\} \cdot p_i \right. \\
\nonumber &\left. - \min \left\{x_i,\theta_i (1-F(p_i)) \right\} \cdot c_i - z_i\cdot s\right]  \\
\nonumber \text{s.t.} &\left\{\delta_i,x_i,y_{ij},z_i,r_{ij} \right\}_{i,j=1}^n \text{ is an equilibrium}\\
  \label{eq:opt_d0} &\text{ under } \{p_i,c_i\}_{i=1}^n\text{ for HV priority assignment}.
\end{align}
The optimization problem \eqref{eq:opt_d0} is difficult to analyze directly. Instead, we propose an equivalent optimization problem, followed by a lemma establishing the equivalence. To this end, consider as an alternative
\begingroup
\allowdisplaybreaks
\begin{align}
\nonumber \max_{\{p_i,\delta_i,x_i,y_{ij},z_i,r_{ij}\}}&\sum_{i=1}^n p_i \theta_i (1-F(p_i)) - \omega\sum_{i=1}^n \delta_i - s\sum_{i=1}^n z_i\\
\nonumber \text{s.t.}\quad d_i=&\theta_i (1-F(p_i))  \\
\nonumber  x_i=&\beta \left[\sum_{j=1}^n \alpha_{ji} \min \left\{ x_j , d_j \right\} + \sum_{j=1}^n y_{ji} \right] + \delta_i\\
\nonumber \sum_{j=1}^n y_{ij}=& \max\left\{ x_i - d_i, 0  \right\}\\
\nonumber  z_i=&\sum_{j=1}^n \alpha_{ji} \max \left\{d_j-x_j , 0\right\} + \sum_{j=1}^n r_{ji}\\
\nonumber \sum_{j=1}^n r_{ij}=&z_i -\max \left\{d_i-x_i , 0\right\}  \\
\label{eq:opt_d1} p_i,&\delta_i,z_i,x_i,y_{ij},r_{ij}\geq 0 \qquad \forall i,j.
\end{align}
\endgroup
In a certain sense formalized in the next lemma, \eqref{eq:opt_d1} is equivalent to \eqref{eq:opt_d0}.

\begin{lemma} 
\label{lem:opt_d021}
\change{Assume HV priority assignment and consider the optimization problems \eqref{eq:opt_d0} and \eqref{eq:opt_d1}. Under Assumption 1, an optimal solution to \eqref{eq:opt_d1} provides an optimal solution to \eqref{eq:opt_d0}. In particular, the following hold:
\begin{enumerate}
    \item If $(1-\beta)\omega< \bar{p}$ or $s<\bar{p}$, then any optimal solution $\left\{p_i^*, \delta^*_i, x_i^*, y_{ij}^*, z_i^*, r_{ij}^* \right\}_{i,j=1}^n$ for \eqref{eq:opt_d1} is such that $d_i^*>0$ for all $i$, \emph{i.e.}, some riders are served at all locations. In this case, there exist compensations $\left\{c^*_i \right\}_{i=1}^n$ such that  $\left\{\delta^*_i, x^*_i, y^*_{ij}, z^*_i, r^*_{ij} \right\}_{i,j=1}^n$ constitutes an equilibrium under $\left\{p^*_i, c^*_i \right\}_{i=1}^n$ for HV priority assignment. Moreover, $\left\{p^*_i, c^*_i \right\}_{i=1}^n$ is optimal for \eqref{eq:opt_d0}.
    \item Conversely, if $(1-\beta)\omega\geq \bar{p}$ and $s\geq\bar{p}$, then any optimal solution for \eqref{eq:opt_d1} and any optimal equilibrium from \eqref{eq:opt_d0} is such that $\delta_i^*=d_i^*=x_i^*=z_i^*=0$ for all $i$, \emph{i.e.}, no riders are served.
\end{enumerate}
}
\end{lemma}

\begin{proofsketchcomment}
The proof of Lemma \ref{lem:opt_d021} closely follows that of \cite[Lemma 1]{Bimpikis2018}, where we adjust the claim and the proof so that it applies to the mixed autonomy setting here. \change{In particular, we first show that the optimal value of \eqref{eq:opt_d1} upper bounds the optimal value of \eqref{eq:opt_d0}. We then show that any feasible solution of \eqref{eq:opt_d1} can be supported as an equilibrium in \eqref{eq:opt_d0} for appropriate choice of compensations.}

\end{proofsketchcomment}

\begin{proofcomment}
\begin{proof}
The proof of the lemma closely follows that of \cite[Lemma 1]{Bimpikis2018}, where we adjust the claim and the proof so that it applies to the mixed autonomy setting here. 

we first show that the optimal value of \eqref{eq:opt_d1} upper bounds the optimal value of \eqref{eq:opt_d0}. For this, we need to show that any solution for \eqref{eq:opt_d0} satisfies $d_i=\theta_i (1-F(p_i)) \leq x_i+z_i$. By contradiction, suppose $d_i>x_i+z_i$, so that increasing the price $p_i$ by a small amount (and thus decreasing $\theta_i (1-F(p_i))$) will improve the value of the objective function. Therefore, $d_i\leq x_i+z_i$ at optimum. Hence we can write the first summation of \eqref{eq:opt_d0} as 
\begin{align}
  \label{eq:10}
  \sum_{i=1}^n \min \left\{x_i+z_i, \theta_i (1-F(p_i)) \right\} =\sum_{i=1}^n \theta_i (1-F(p_i)).
\end{align}
The term $\omega\sum_i\delta_i$ is the cost rate for drivers of the platform, which is a lower bound for the platform's cost on human-driven vehicles at equilibrium. Moreover, the constraints in \eqref{eq:opt_d1} correspond to the equilibrium constraints in \eqref{eq:opt_d0}. Therefore, the optimal value of \eqref{eq:opt_d1} is an upper bound for that of \eqref{eq:opt_d0}.

Next, we'll see that the upper bound can be reached by the optimal solution supported by some compensations $\left\{c_i \right\}_{i=1}^n$ under equilibrium.

To prove the second part of the lemma, we construct a compensation $\left\{c_i \right\}_{i=1}^n$ so that $V_i = \omega $ for all $i$. To that end, let 
\begin{align}
  \label{eq:11}
  c_i &= \begin{cases}
                  \frac{x_i}{d_i}\cdot\omega(1-\beta) &\text{if }d_i<x_i\\
                  \omega(1-\beta) & \text{if } d_i\geq x_i.
\end{cases}
\end{align}

 Since we assumed that $d_i > 0$ for all $i$, then $c_i < \infty$ for all $i$ and thus the compensation is well-defined. Moreover, the probability that any driver at location $i$ is assigned to a ride is $\frac{d_i}{x_i}$ when $d_i< x_i$ and is 1 when $d_i \geq x_i$ since the driver takes the priority when drivers and AVs both exist in the platform. Therefore, the expected earnings for a single time period for a driver at location $i$ are equal to $\omega(1-\beta)$. Thus, the expected lifetime earnings are $V_i = \sum_j^\infty \beta^j\omega(1-\beta) = \omega$. Hence, the solution $\left\{p_i, \delta_i, x_i, y_{ij}, z_i, r_{ij} \right\}_{i,j=1}^n$ is supported as an equilibrium using the compensations we constructed above.

Moreover, the cost incurred by the platform under these compensations per period is
\begin{equation*}
\sum_{i=1}^n \min \left\{x_i,\theta_i (1-F(p_i)) \right\}\cdot c_i = \sum_{i=1}^n \min \left\{x_i,d_i \right\} \cdot c_i.
\end{equation*}
We construct a partition for the locations so that $I_1 = \left\{i: d_i<x_i \right\}$ and $I_2 = \left\{i: d_i\geq x_i \right\}$. Therefore 
\begin{align*}
 \sum_i \min \left\{x_i,d_i \right\} \cdot c_i =& \sum_{i\in I_1} d_i c_i + \sum_{i\in I_2} x_i c_i \\
 =& \sum_{i\in I_1} d_i\cdot \frac{x_i}{d_i}\omega(1-\beta) + \sum_{i\in I_2} x_i \omega(1-\beta)\\
 =& \sum_i x_i\omega(1-\beta) = \sum_{i=1}^n \delta_i\omega.
\end{align*}
The last equality follows from the fact that $\sum_{i=1}^n x_i(1-\beta) = \sum_{i=1}^n\delta_i$ since, at equilibrium, the mass of drivers entering the platform is equal to the mass of drivers that are leaving.

The third part of the lemma follows directly from the second part of \cite[Lemma 1]{Bimpikis2018} since $z_i>0$ only if $d_i>0$ in our scenario.
\end{proof}
\end{proofcomment}
Turning now to the case of AV priority assignment, the analogous profit-maximization problem is given by \eqref{eq:opt_a0} below and as in the case of HV priority assignment, we introduce \eqref{eq:opt_a1} for AV priority assignment.
\begin{align}
\nonumber \max_{\{p_i,c_i\}_{i=1}^n}&\sum_{i=1}^n \left[ \min \left\{x_i+z_i, \theta_i (1-F(p_i)) \right\} \cdot p_i \right. \\
\nonumber &\left.  - \min \{x_i, \max \{ \theta_i(1-F(p_i))-z_i,0\} \} \cdot c_i - z_i\cdot s\right]  \\
\nonumber \text{s.t.} \{&\delta_i,x_i,y_{ij},z_i,r_{ij} \}_{i,j=1}^n \text{ is an equilibrium under}\\
\label{eq:opt_a0} &	\{p_i,c_i\}_{i=1}^n\text{ for AV priority assignment}.
\end{align}
\begin{align}
\nonumber \max_{\{p_i,\delta_i,x_i,y_{ij},z_i,r_{ij}\}}&\sum_{i=1}^n p_i \theta_i (1-F(p_i)) - \omega\sum_{i=1}^n \delta_i - s\sum_{i=1}^n z_i\\
\nonumber \text{s.t.}\quad d_i=&\theta_i (1-F(p_i))  \\
\nonumber  x_i=&\beta \left[\sum_{j} \alpha_{ji} \max \left\{ d_j-z_j,0\right\} + \sum_{j} y_{ji} \right] + \delta_i\\
\nonumber \sum_{j=1}^n y_{ij}=& x_i -\max \left\{ d_i-z_i,0\right\}  \\
\nonumber  z_i=&\sum_{j=1}^n \alpha_{ji} \min \left\{d_j, z_j\right\} + \sum_{j=1}^n r_{ji}\\
\nonumber \sum_{j=1}^n r_{ij}=& \max \left\{z_i-d_i , 0\right\}  \\
  \label{eq:opt_a1} p_i,&\delta_i,z_i,x_i,y_{ij},r_{ij}\geq 0 \qquad \forall i,j.
\end{align}
Mirroring Lemma \ref{lem:opt_d021}, optimization problems \eqref{eq:opt_a0} and \eqref{eq:opt_a1} are equivalent in a certain sense.

\begin{lemma} 
\label{lem:opt_a021}
\change{Assume AV priority assignment and consider the optimization problems \eqref{eq:opt_a0} and \eqref{eq:opt_a1}. Under Assumption \ref{assum:all}, an optimal solution to \eqref{eq:opt_a1} provides an optimal solution to \eqref{eq:opt_a0}. In particular, the following hold:
\begin{enumerate}
    \item If $(1-\beta)\omega< \bar{p}$ or $s<\bar{p}$, then any optimal solution $\left\{p_i^*, \delta^*_i, x_i^*, y_{ij}^*, z_i^*, r_{ij}^* \right\}_{i,j=1}^n$ for \eqref{eq:opt_a1} is such that $d_i^*>0$ for all $i$, \emph{i.e.}, some riders are served at all locations. In this case, there exist compensations $\left\{c^*_i \right\}_{i=1}^n$ such that $\left\{\delta^*_i, x^*_i, y^*_{ij}, z^*_i, r^*_{ij} \right\}_{i,j=1}^n$ constitutes an  equilibrium under $\left\{p^*_i, c^*_i \right\}_{i=1}^n$ for AV priority assignment. Moreover, $\left\{p^*_i, c^*_i \right\}_{i=1}^n$ is optimal for \eqref{eq:opt_a0}. 
    \item Conversely, if $(1-\beta)\omega\geq \bar{p}$ and $s \geq \bar{p}$, then any optimal solution for \eqref{eq:opt_a1} and any optimal equilibrium from \eqref{eq:opt_a0}  is such that $\delta_i^*=d_i^*=x_i^*=z_i^*=0$ for all $i$, \emph{i.e.}, no riders are served.
   \end{enumerate}
}
\end{lemma}

\begin{proofsketchcomment}
The proof is similar to that of Lemma \ref{lem:opt_d021}.
\change{
}
\end{proofsketchcomment}

\begin{proofcomment}
The proof is similar to that of Lemma \ref{lem:opt_d021} by setting 
\begin{align*}
  c_i &= \begin{cases}
                  \frac{x_i}{d_i-z_i}\cdot\omega(1-\beta) &\text{if }d_i>z_i\\
                  \omega(1-\beta) & \text{if } d_i\leq z_i.
\end{cases}
\end{align*}
\end{proofcomment}

From Lemma \ref{lem:opt_d021} (resp., Lemma \ref{lem:opt_a021}), we conclude that it is without loss of generality for us to focus on the optimization problem \eqref{eq:opt_d1} (resp., \eqref{eq:opt_a1}) for the rest of the paper when considering HV (resp., AV) priority assignment.

\change{Moreover, while the objective function of \eqref{eq:opt_d1} (resp., \eqref{eq:opt_a1}) is not concave in general, it is concave for distributions for which the term $p\cdot(1-F(p))$ is concave in the fractional demand  $d=1-F(p)$, which can be set by the platform by adjusting the price $p$ (note that $p\cdot d= d\cdot F^{-1}(1-d)$). For example, the uniform distribution, exponential distribution and Pareto distribution all satisfy this concavity requirement. Throughout the rest of the paper, we focus on the case where the rider's willingness to pay is such that the revenue of the platform is concave in $d$.}

\change{
\begin{assum}
\label{assum:obj_concave}
The cumulative distribution $F(\cdot)$ of the riders' willingness to pay is such that $d\cdot F^{-1}(1-d)$ is concave in $d$.
\end{assum}
}

Under HV (resp., AV) priority assignment, we have converted \eqref{eq:opt_d0} (resp., \eqref{eq:opt_a0}) to the alternative optimization problem \eqref{eq:opt_d1} (resp., \eqref{eq:opt_a1}). Next, we will further convert \eqref{eq:opt_d1} (resp., \eqref{eq:opt_a1}, henceforth written as \eqref{eq:opt_d1}/\eqref{eq:opt_a1}) to an alternative optimization problem that is also convex, allowing for efficient---and in some cases, closed form---solution computation.

\section{Convexification of Profit Maximization}
\label{sec:alt-orig-optim}
Even when \eqref{eq:opt_d1}/\eqref{eq:opt_a1} possesses a concave objective function, the constraints are non-convex \change{and cannot be simply convexified} so that solving \eqref{eq:opt_d1}/\eqref{eq:opt_a1} remains computationally difficult, \emph{i.e.}, nonconvex. This section introduces alternative optimization problems of the mixed autonomy deployment for which the optimal profits will be the same as that of \eqref{eq:opt_d1}/\eqref{eq:opt_a1}.

While the optimal profits are the same, the optimal solutions of the alternative optimization problems are not exactly the same as those calculated in the original problems \eqref{eq:opt_d1}/\eqref{eq:opt_a1}. \change{As a result, a main difference between the original problems and their alternatives is that, while the original problems and their optimal solutions can always be interpreted physically, the alternatives are purely mathematical problems.} However, given the optimal solution of the alternative problems, we show that it is possible to compute an optimal solution for the original problems \eqref{eq:opt_d1}/\eqref{eq:opt_a1} with identical profit and vice versa. \change{Moreover, by eliminating $p_i$ using $d_i = \theta_i(1-F(p_i))$ in substitution, the alternative optimization problems are seen to be convex optimization problems under Assumption \ref{assum:obj_concave}. But, for clarity, we leave  $p_i$ in the alternative optimization problems to allow for comparison to the original problems. Furthermore, the alternative optimization problems}
become quadratic optimization problems with linear constraints when $F(\cdot)$ is a uniform distribution.

First, assume HV priority assignment, and consider the  optimization problem given by
\begin{align}
\nonumber \max_{\{p_i,\delta_i,x_i,z_i,r_{ij}\}}&\sum_{i=1}^n p_i \theta_i (1-F(p_i)) - \omega\sum_{i=1}^n \delta_i - s\sum_{i=1}^n z_i\\  
\nonumber  \text{s.t.}\quad d_i&=\theta_i (1-F(p_i))\\
\nonumber  x_i&=\beta \sum_{j=1}^n \alpha_{ji}x_j + \delta_i\\
\nonumber  z_i&=\sum_{j=1}^n \alpha_{ji} (d_j-x_j)+ \sum_{j=1}^n r_{ji}\\
\nonumber \sum_{j=1}^n r_{ij}&=z_i -(d_i-x_i )\\
  \label{eq:opt_d2} p_i,&\delta_i,x_i,z_i,r_{ij}\geq 0 \qquad \forall i,j.
\end{align}

In the following, we regard \eqref{eq:opt_d1} as the \emph{original} optimization problem and \eqref{eq:opt_d2} as the \emph{alternative} optimization problem for HV priority assignment.

Theorem \ref{thm:equiv_driver} below states that \eqref{eq:opt_d1} and \eqref{eq:opt_d2} have the same optimal profits for any $\beta$, $s$, $\omega$ and adjacency matrix $\mathbf{A}$. Moreover, given one optimal solution for \eqref{eq:opt_d1} or \eqref{eq:opt_d2}, it is possible to compute an optimal solution for the other.

\begin{thm}
\label{thm:equiv_driver}
Assume HV priority assignment, and consider the original optimization problem \eqref{eq:opt_d1} and alternative optimization problem \eqref{eq:opt_d2}. Let
\begin{align}
  \label{eq:d-ori-sol}
  \mathbf{u}^{ori*} = \left\{p_i^{ori*}, \delta_i^{ori*}, z_i^{ori*}, x_i^{ori*}, y_{ij}^{ori*}, r_{ij}^{ori*}\right\}_{i,j=1}^n
\end{align}
be an optimal solution for \eqref{eq:opt_d1} and 
\begin{align}
  \label{eq:d-alt-sol}
  \mathbf{u}^{alt*} = \left\{p_i^{alt*}, \delta_i^{alt*}, z_i^{alt*}, x_i^{alt*}, r_{ij}^{alt*}\right\}_{i,j=1}^n
\end{align}
be an optimal solution for \eqref{eq:opt_d2}. Then the following hold under Assumptions \ref{assum:all} and \ref{assum:obj_concave}:
\begin{itemize}
\item The original optimization problem and the alternative problem obtain the same optimal profits for all possible choices of $\beta$, $s$, $\omega$ and adjacency matrix $\mathbf{A}$.

\item The optimal solutions satisfy $x^{ori*} =  x^{alt*}$, $z^{ori*} = z^{alt*}$, $p^{ori*} = p^{alt*}$ and $\delta^{ori*} = \delta^{alt*}$.

\item If $\theta_i (1-F(p^{ori*}_i))\leq x_i^{ori*}$ for all $i$ in the original optimization problem, then $z_i^{ori*}=0$ for all $i$ and setting $r^{alt*}_{ij} = y^{ori*}_{ij}$ for all $i,j$ constitutes an optimal solution for the alternative problem. 

\item If $ \theta_i (1-F(p^{alt*}_i))\leq x_i^{alt*}$ for all $i$ in the alternative optimization problem, then $z_i^{alt*}=0$ for all $i$ and setting $y^{ori*}_{ij} = r^{alt*}_{ij}$, $r^{ori*}_{ij} = 0$ constitutes an optimal solution for the original optimization problem. 
\end{itemize}
\end{thm}
\begin{proofsketchcomment}
\begin{proofsketch}
Let $\profit^{ori*}$ and $\profit^{alt*}$ be the optimal profits of the two problems \eqref{eq:opt_d1} and \eqref{eq:opt_d2}, respectively, and let $d^{ori*}_i=\theta_i (1-F(p^{ori*}_i))$ and $d^{alt*}_i= \theta_i (1-F(p^{alt*}_i))$. To prove that the optimal profits of the two problems are equal, we first show that $\profit^{ori*} \leq  \profit^{alt*}$ and then $\profit^{ori*} \geq  \profit^{alt*}$.

To prove the claim in the direction $\profit^{ori*} \leq  \profit^{alt*}$, we consider three cases: $d^{ori*}_i\geq x^{ori*}_i$ for all $i$, $d^{ori*}_i\leq x^{ori*}_i$ for all $i$, and the heterogeneous case when there exists some location $i$ such that $x^{ori*}_i > d^{ori*}_i$ and some location $j$ such that $x^{ori*}_j < d^{ori*}_j$. Similarly, in the opposite direction establishing $\profit^{ori*} \geq  \profit^{alt*}$, we consider three cases: $d^{alt*}_i\geq x^{alt*}_i$ for all $i$, $d^{alt*}_i\leq x^{alt*}_i$ for all $i$, and the heterogeneous case when $x^{alt*}_i > d^{alt*}_i$ for some location $i$ and $x^{alt*}_j < d^{alt*}_j$ for some location $j$.

For each direction, if the optimal solution of \eqref{eq:opt_d1}/\eqref{eq:opt_d2} falls in the first two cases, we prove the claim by showing the existence of a feasible solution for \eqref{eq:opt_d2}/\eqref{eq:opt_d1} with the same profit. 

We next show that the heterogeneous case in the first direction, for which there exists some location $i$ such that $x^{ori*}_i > d^{ori*}_i$ and some location $j$ such that $x^{ori*}_j < d^{ori*}_j$, is not possible. \change{We consider an aggregated network by partitioning all locations into two or three combined nodes. Then we illustrate that this heterogeneous situation contradicts the Assumption \ref{assum:all} which requires the network to be strongly connected. }
In the other direction, we make use of the KKT conditions, which narrows the range of possible relations between $s$ and $\omega$, in order to establish again the existence of a feasible solution for \eqref{eq:opt_d2}/\eqref{eq:opt_d1} with the same profit, completing the proof.
\end{proofsketch}
\end{proofsketchcomment}

\begin{proofcomment}
\begin{proof}

Let $\profit^{ori*}$ and $\profit^{alt*}$ be the optimal profits of the two problems \eqref{eq:opt_d1} and \eqref{eq:opt_d2}, respectively, and let $d^{ori*}_i=\theta_i (1-F(p^{ori*}_i))$ and $d^{alt*}_i= \theta_i (1-F(p^{alt*}_i))$.

To prove that the optimal profits of the two problems are equal, we first show that $\profit^{ori*} \leq  \profit^{alt*}$ and then $\profit^{ori*} \geq  \profit^{alt*}$.

We first introduce Lagrange multiplies $\lambda_i$, $\mu_i$, and $\gamma_i$ and establish the following inequalities for all $i,j$ derived from the KKT conditions that are necessary for any optimal solution of \eqref{eq:opt_d2}:
\begin{align}
  \label{eq:dk-1} &\text{(constraints on $\delta_i$)}&&-\omega +\lambda_i \leq 0\\ 
  \label{eq:dk-2} 
  &\text{(constraints on $x_i$)} &&\sum_j \alpha_{ij} (\beta \lambda_j-\mu_j)-\lambda_i+\gamma_i\leq 0 \\
  \label{eq:dk-3} 
  &\text{(constraints on $z_i$)} &&-s + \gamma_i-\mu_i\leq 0\\
  \label{eq:dk-4}
  &\text{(constraints on $r_{ij}$)} &&\mu_j-\gamma_i \leq 0.
\end{align}
We now consider three cases to prove $\profit^{ori*} \leq  \profit^{alt*}$.

\underline{Case 1:} $d^{ori*}_i\geq x^{ori*}_i$ for all $i$. Then $\mathbf{u}^{ori*}$ is feasible for the alternative problem because both problems are in fact the same optimization problem in this case. Therefore $\profit^{ori*} \leq  \profit^{alt*}$.

\underline{Case 2:} $d^{ori*}_i\leq x^{ori*}_i$ for all $i$. Then the AVs are not needed in any location and $z_i = 0, r_{ij} = 0 \quad \forall i,j$. Then the original optimization problem becomes
\begin{align}
\nonumber \max_{\{p_i,\delta_i,x_i,y_{ij},z_i,r_{ij}\}} \qquad &\hspace{-5pt}\sum_{i=1}^n p_i \theta_i (1-F(p_i)) - \omega\sum_{i=1}^n \delta_i\\
\nonumber  \text{s.t.}\quad d_i&=\theta_i (1-F(p_i))\\
\nonumber  x_i&=\beta \left[\sum_{j=1}^n \alpha_{ji} d_j + \sum_{j=1}^n y_{ji} \right] + \delta_i\\
\nonumber \sum_{j=1}^n y_{ij}&= x_i - d_i\\
  \label{eq:d2_1} p_i,\delta_i,x_i,y_{ij}&\geq 0 \qquad \forall i,j.
\end{align}
Let $z^{alt}_i=0$ and $y^{alt}_{ij}=0 \quad \forall i,j$. Then the alternative problem becomes exactly the same problem as \eqref{eq:d2_1} when we substitute $r_{ij}$ with $y_{ij}$, which proves the claim.

\underline{Case 3:} There exists some location $i$ such that $x^{ori*}_i > d^{ori*}_i$ and some location $j$ such that $x^{ori*}_j < d^{ori*}_j$. In this case, if there is no $i$ such that  $x^{ori*}_i = d^{ori*}_i$, then let $I_1=\{i: x^{ori*}_i > d^{ori*}_i \}$ and let $I_2=\{i: x^{ori*}_i < d^{ori*}_i \}$. We can then consider an aggregated network with locations $1$ and $2$ representing the combined locations in $I_1$ and $I_2$, respectively. 

Hence, in this aggregated network, $\alpha_{11} = \alpha_{22} \geq 0$; $\alpha_{12} >0$ and $\alpha_{21} >0$ by our assumption that the directed graph defined by adjacency matrix $\mathbf{A}$ is strongly connected.

Since $d^{ori*}_1<x^{ori*}_1$, then $z^{ori*}_1 = 0$. On the other hand, $z^{ori*}_1 = \max \left\{d^{ori*}_1-x^{ori*}_1 , 0\right\} \alpha_{11} + \max \left\{d^{ori*}_2-x^{ori*}_2 , 0\right\} \alpha_{21} + \sum_{j=1}^2 r^{ori*}_{j1} = (d^{ori*}_2-x^{ori*}_2)\alpha_{21}+\sum_{j=1}^2 r^{ori*}_{j1}$ since $d^{ori*}_2-x^{ori*}_2 >0$ and $d^{ori*}_1-x^{ori*}_1 <0$. Hence $z_1^{ori*}>0$ which leads to a contradiction. Therefore, if there is no $i$ such that  $x^{ori*}_i = d^{ori*}_i$, then either $x^{ori*}_i > d^{ori*}_i \quad$ for all $i$ or $x^{ori*}_i < d^{ori*}_i$  for all $i$.

If there exists $i$ such that $x^{ori*}_i = d^{ori*}_i$, define $I_1$ and $I_2$ as above and introduce $I_3=\{i: x^{ori*}_i = d^{ori*}_i\}$. 

Similar to the above argument, we show that $z^{ori*}_1 =z^{ori*}_3 = 0$. Since $z^{ori*}_1=\sum_{j=1}^3 \alpha_{j1} \max \left\{d^{ori*}_j-x^{ori*}_j , 0\right\} + \sum_{j=1}^3 r^{ori*}_{j1}$ while $d^{ori*}_2-x^{ori*}_2>0$, then $\alpha_{21}=0$. Similarly, we must have $\alpha_{23}=0$. Therefore, we have $\alpha_{22}=1$ since $\sum_{j=1}^3 \alpha_{ij}=1$. However, $\alpha_{22}=1$ means that some components in the graph are not strongly connected with the others, which contradicts our assumption. Hence this mixed situation cannot be an optimal solution for the problem.

Thus, up to now, we have shown that $\profit^{ori*} \leq  \profit^{alt*}$.
Next we show that $\profit^{ori*} \geq  \profit^{alt*}$. 

\underline{Case 1:} If $d^{alt*}_i\geq x^{alt*}_i$ for all $i$, then $\mathbf{u}^{ori*}$ is feasible for the original problem because both problems are in fact the same optimization problem in this case. Therefore $\profit^{ori*} \geq  \profit^{alt*}$.

\underline{Case 2:} If $d^{alt*}_i\leq x^{alt*}_i$ for all $i$, we want to show that in this case, $z^{alt*}_i=0$ for all $i$ and then $\mathbf{u}^{alt*}$ will be feasible for the original optimization by setting $y^{ori}_{ij} = r^{alt}_{ij}$ with $r^{ori}_{ij}=0$ for all $i,j$. 

	Fix $d^{alt*}_i\leq x^{alt*}_i$ for all $i$, then if $z_i=0$ is a feasible solution for \eqref{eq:opt_d2}, then it will be the optimal the solution since any increase in $z_i$ will increase the cost and reduce the profit. 
	
	We'll show below that given $d^{alt*}_i\leq x^{alt*}_i$ and setting $z_i=0$ for all $i$ for \eqref{eq:opt_d2}, there exists $r_{ij}$ that satisfies the constraints for \eqref{eq:opt_d2} and thus constitutes a feasible solution for the alternative optimization problem.

\begin{align}
\nonumber \sum_{j=1}^n r_{ij}&= x_i - d_i\\
\nonumber \sum_{j=1}^n r_{ji}&= \sum_{j=1}^n \alpha_{ji} (x_j - d_j)\\
 \label{eq:d_t1} r_{ij}, (x_i - d_i) &\geq 0 \qquad \forall i,j.
\end{align}
	
The new constraints can be described as in \eqref{eq:d_t1}. We can reformulate \eqref{eq:d_t1} into \eqref{eq:d_t2} below where $\mathbf{R}$ is an $n$ by $n$ matrix and  $[ \mathbf{R} ]_{ij} = r_{ij}$; $\mathbf{\Delta}$ is an $n$ by $1$ vector and  $[ \mathbf{\Delta} ]_{i} = x_i - d_i$; $\mathbf{1}$ is an $n$ by $1$ one's vector.
	
\begin{align}
\nonumber  \mathbf{R}\mathbf{1} = \mathbf{\Delta}\\
\nonumber  \mathbf{R}^T\mathbf{1} =A^T \mathbf{\Delta}\\
\nonumber  \mathbf{\Delta}\geq0\\
 \label{eq:d_t2} \mathbf{R}_{ij} \geq 0
 \end{align}
 
We can then vectorize $\mathbf{R}$ to $\mathbf{\hat{R}}$ (in row) so that \eqref{eq:d_t2} will transform into \eqref{eq:d_t3}.
	
\begin{align}
\nonumber  \mathbf{M}\mathbf{\hat{R}} = \mathbf{b}\\
 \label{eq:d_t3} \mathbf{\hat{R}}_{ij} \geq 0
 \end{align}
 	 $\mathbf{M} = \begin{bmatrix}  \mathbf{M}_1\\ \mathbf{M}_2 \end{bmatrix}$ where $\mathbf{M}_1$ and $\mathbf{M}_2$ are both $n$ by $n^2$ matrices:

$\mathbf{M}_1 = \mathbf{I}_{n\times n}\otimes \mathbf{1}^T =  \begin{bmatrix}  1 & \hdots & 1 &  0 & \hdots & 0 & \hdots & 0 &  \hdots & 0 \\
	  0 & \hdots & 0 & 1 & \hdots & 1 & \hdots & 0 & \hdots & 0 \\
	    &  \vdots&  &  & \vdots  &  &  \vdots &  &  \vdots & \\ 
	  0 & \hdots & 0 & 0 & \hdots & 0 & \hdots & 1 & \hdots & 1  \end{bmatrix}$
	  and	  
	  
$\mathbf{M}_2 = \mathbf{1}^T \otimes \mathbf{I}_{n\times n}  =  \begin{bmatrix}  \mathbf{I}_{n\times n} & \mathbf{I}_{n\times n} & \hdots &  \mathbf{I}_{n\times n}  \end{bmatrix}$.
	  
	 $\mathbf{\hat{R}} = [\mathbf{R}_{11},\mathbf{R}_{12},\hdots,\mathbf{R}_{1n},\hdots,\mathbf{R}_{n1},\mathbf{R}_{n2},\hdots,\mathbf{R}_{nn}]^T$ is a $n^2$ by $1$ vector.
	 
	  $ \mathbf{b} = \begin{bmatrix} \mathbf{\Delta} \\ A^T \mathbf{\Delta} \end{bmatrix}$ is a $2n$ by $1$ vector.
	  
By Farka's Lemma, to prove that \eqref{eq:d_t3} has a feasible solution $\mathbf{\hat{R}}$: that is, $\exists  \mathbf{\hat{R}}$ s.t. $\mathbf{M}\mathbf{\hat{R}} = \mathbf{b}$ and $\mathbf{\hat{R}} \geq 0$, we only need to disprove the claim that $\exists  \mathbf{v} \in \mathbb{R}^{2n}$ s.t. $\mathbf{M}^T \mathbf{v}\geq 0$ and $ \mathbf{b}^T \mathbf{v} <0$. Denote $v_i$ as the $i$th element of $\mathbf{v}$.
	  
	 Let $\mathbf{v} \in \mathbb{R}^{2n}$ s.t. $\mathbf{M}^T \mathbf{v}\geq 0$, 
\begin{align*}
\mathbf{M}^T \mathbf{v} & = \begin{bmatrix} \mathbf{M}_1^T & \mathbf{M}_2^T \end{bmatrix} \mathbf{v} = \begin{bmatrix} 
	 \mathbf{1} & \mathbf{0} & \hdots & \mathbf{0} & \mathbf{I}_{n\times n} \\
	   \mathbf{0} & \mathbf{1} & \hdots & \mathbf{0} & \mathbf{I}_{n\times n} \\
	    \mathbf{0} & \mathbf{0} & \hdots & \mathbf{1} & \mathbf{I}_{n\times n}  \end{bmatrix}\mathbf{v}.
\end{align*}
Hence, $v_i + v_j \geq 0$ for all $i = 1,\hdots,n$ and $j = n+1,\hdots,2n$.
	     
Now consider $ \mathbf{b}^T \mathbf{v}$. 
\begin{align*}	
\mathbf{b}^T \mathbf{v} &= \begin{bmatrix} \mathbf{\Delta}^T & \mathbf{\Delta}^T A \end{bmatrix} \mathbf{v} = \mathbf{\Delta}^T \begin{bmatrix} \mathbf{I}_{n\times n} & A  \end{bmatrix} \mathbf{v}  \\
	&= \mathbf{\Delta}^T \begin{bmatrix} \vdots \\ v_i+\sum_{j=1}^n \alpha_{ij} v_{j+n} \\ \vdots \end{bmatrix}\\
	& =  \mathbf{\Delta}^T \begin{bmatrix} \vdots \\ \sum_{j=1}^n \alpha_{ij}(v_i+ v_{j+n}) \\ \vdots \end{bmatrix}     
\end{align*}
The last equality comes from the fact that $\sum_{j=1}^n \alpha_{ij} = 1$. Moreover, since $v_i+ v_{j+n}\geq 0$ for all $i = 1,\hdots,n$ as previously mentioned, and $\mathbf{\Delta}\geq 0$, then $\mathbf{b}^T \mathbf{v} \geq 0$. Hence we disproved the claim that  $\exists  \mathbf{v} \in \mathbb{R}^{2n}$ s.t. $\mathbf{M}^T \mathbf{v}\geq 0$ and $ \mathbf{b}^T \mathbf{v} <0$.
	     
Therefore \eqref{eq:d_t3} has a feasible solution $\mathbf{\hat{R}}$ and thus \eqref{eq:d_t1} has feasible solution $r_{ij}$ for all $i,j$. Hence $z^{alt*}_i=0$ for all $i$ and then $\mathbf{u}^{alt*}$ will be feasible for the original optimization by setting $y^{ori}_{ij} = r^{alt}_{ij}$ with $r^{ori}_{ij}=0$ for all $i,j$. 

\underline{Case 3:}  There exist $\beta$ and $k$ such that the optimal solution $\mathbf{u}^{alt*}$ does not satisfy the two situations above, which means there exist locations such that $d^{alt*}_i > x^{alt*}_i$ for some $i$ and $d^{alt*}_j < x^{alt*}_j$ for some $j$.  Let $I_1=\{i: x^{alt*}_i < d^{alt*}_i \}$ and let $I_2=\{i: x^{alt*}_i \geq d^{alt*}_i \}$ and we can consider an aggregated network with locations $1$ and $2$ representing the combined locations in $I_1$ and $I_2$, respectively. Knowing $x^{alt*}_1 < d^{alt*}_1$, suppose $x^{alt*}_2 > d^{alt*}_2$ (since there exists at least an $i$ such that $d^{alt*}_i < x^{alt*}_i$). Then we can rewrite the constraints of \eqref{eq:opt_d2} as below:

\begin{align}
\nonumber x_1 = &\beta(\alpha_{11}x_1 +\alpha_{21}x_2  ) +\delta_1\\
\nonumber x_2 = &\beta(\alpha_{12}x_1 +\alpha_{22}x_2) + \delta_2 \\
\nonumber  z_1=&\alpha_{11}(d_1-x_1)+\alpha_{21}(d_2-x_2)+ r_{11}+ r_{21}\\
\nonumber  z_2=&\alpha_{12}(d_1-x_1)+\alpha_{22}(d_2-x_2)+ r_{12}+ r_{22}\\
\nonumber r_{11} +r_{12}=& z_1-( d_1-x_1) \\
\nonumber r_{21} +r_{22}=& z_2-( d_2-x_2) \\
  \label{eq:d_t4} p_i,&\delta_i,z_i,x_i,r_{ij}\geq 0 \qquad \forall i,j.
\end{align}

For convenience, denote $\Delta_1 =d^{alt*}_1-x^{alt*}_1 $ and $\Delta_2 =d^{alt*}_2-x^{alt*}_2 $. Obviously, $\Delta_1 >0 $ and $\Delta_2<0$. 

Since $r_{11} +r_{12}\geq 0$, then $z^{alt*}_1>\Delta_1>0$ and this indicates that $\gamma_1-\mu_1 = s$. Hence $\mu_1-\gamma_1 = -s \neq 0$ and thus $r^{alt*}_{11} =0$.

Since $x^{alt*}_2 > d^{alt*}_2 > 0$, then $x^{alt*}_1>0$ since $\alpha_{21}>0$ for strong connectivity of the network. Moreover, these indicates that $\delta^{alt*}_1+\delta^{alt*}_2 = (1-\beta)(x^{alt*}_1+x^{alt*}_2)>0$

As $z_2\geq0$, then $r^{alt*}_{21} +r^{alt*}_{22}\geq x^{alt*}_2 - d^{alt*}_2>0$. 
Suppose $r^{alt*}_{21} = 0$, then $r^{alt*}_{22} > 0$, then $\mu_2-\gamma_2 = 0$ and hence $z^{alt*}_2 = 0$. Then $z^{alt*}_1 = \alpha_{11}\Delta_1+\alpha_{21}\Delta_2$. Knowing $z^{alt*}_1 \geq \Delta_1$ requires $\alpha_{11} = 1, \alpha_{21} = 0$ (because $\Delta_2<0$) and this network is no longer strongly connected which contradicts the assumption. Therefore $r^{alt*}_{21} > 0$ and thus $\mu_1 - \gamma_2 =0$. We can get $\mu_2 - \gamma_1 = (\mu_2 - \gamma_2) +(\gamma_2 -\mu_1)+(\mu_1 - \gamma_1) \leq 0+0-s<0$ so that $r^{alt*}_{12}=0$. Therefore $z^{alt*}_1 = \Delta_1$; $r^{alt*}_{21} = z^{alt*}_1 -\alpha_{11}\Delta_1-\alpha_{21}\Delta_2 = \alpha_{12}\Delta_1-\alpha_{21}\Delta_2$.

With all the preliminary results above, we now divide the problem into two cases: $z^{alt*}_2 = 0$ or $z^{alt*}_2 >0$.

Suppose $z^{alt*}_2 = 0$, $r^{alt*}_{22}  = \Delta_2 - r^{alt*}_{21} =  -\alpha_{12}\Delta_1-\alpha_{22}\Delta_2 > 0$, which implies that $\alpha_{12}\Delta_1< -\alpha_{22}\Delta_2$ and $\mu_2 - \gamma_2 = 0$. Hence $\gamma_1 - \mu_2 = (\gamma_1 - \mu_1) +(\mu_1 - \gamma_2)+( \gamma_2 - \mu_2)  = s+0+0 = s$. 

Then \eqref{eq:dk-2} yields that
\begin{align}
\label{d_t5}	\beta( \alpha_{11}\lambda_1+\alpha_{12}\lambda_2) - \lambda_1 + \alpha_{11}s + \alpha_{12}s &= 0\\
\label{d_t6}	\beta( \alpha_{21}\lambda_1+\alpha_{22}\lambda_2) - \lambda_2 + \alpha_{21}\cdot0 + \alpha_{22}\cdot0&= 0 
\end{align}

If $\lambda_2 = \omega$, then \eqref{d_t6} shows that $\beta \alpha_{21}\lambda_1 =(1-\beta\alpha_{22}) \lambda_2 >(\beta-\beta\alpha_{22}) \lambda_2 = \beta \alpha_{21}\lambda_2$. Hence $\lambda_1>\lambda_2>\omega$ which contradicts to \eqref{eq:dk-1}. Therefore, $\lambda_2 < \omega \Rightarrow \delta^{alt*}_2 = 0$. Since $\delta^{alt*}_1+\delta^{alt*}_2>0$, then $\delta^{alt*}_1>0$ and $\lambda_1 = \omega$, $\lambda_2 = \frac{\beta\alpha_{21}}{1-\beta\alpha_{22}}\cdot\omega$. Applying this result to \eqref{d_t5} gives $s = \frac{(1-\beta)(1+\beta\alpha_{12}-\beta\alpha_{22})}{1-\beta\alpha_{22}}\cdot\omega >(1-\beta)\omega$.

Let $p^{ori}_i = p^{alt*}_i$, $r^{ori}_{ij} = 0$, $y^{ori}_{ij} = r^{alt*}_{ij}$ for all $i,j$; $\delta^{ori}_1 = (1-\beta)(d^{alt*}_1+x^{alt*}_2), \delta^{ori}_2=0, z^{ori}_1 = z^{ori}_2 =0$, $x^{ori}_1 = d^{alt*}_1$ and $x^{ori}_2 = x^{alt*}_2$. Then $ \mathbf{u}^{ori} = \left\{p^{ori}_i, \delta^{ori}_i, z^{ori}_i, x^{ori}_i, y^{ori}_{ij}, r^{ori}_{ij}\right\}_{i,j=1}^2$ would be a feasible solution for \eqref{eq:opt_d1}. This solution increases the cost by $\omega \cdot (\delta^{ori}_1- \delta^{alt*}_1+\delta^{ori}_2 -\delta^{alt*}_2) = (1-\beta)\omega \Delta_1$, decreases the cost by $s\cdot(z^{alt*}_1- z^{ori}_1+z^{alt*}_2 -z^{ori}_2) = s\cdot \Delta_1 > (1-\beta)\omega \Delta_1$. The net profit increases, hence there always exists a solution for the original optimization problem that has a higher profit and thus the solution is not optimal (since we've already proved that $\profit^{ori*} \leq  \profit^{alt*}$ ).

Therefore $z^{alt*}_2 >0$, which indicates $r^{alt*}_{22}=0$ and $\mu_2-\gamma_2 = s$. Hence $\gamma_1 - \mu_2 = (\gamma_1 - \mu_1) +(\mu_1 - \gamma_2)+( \gamma_2 - \mu_2)  = s+0+s = 2s$. Moreover, $z^{alt*}_2 = \alpha_{12}\Delta_1+\alpha_{22}\Delta_2>0$ implies $\alpha_{12}\Delta_1> -\alpha_{22}\Delta_2$

Then \eqref{eq:dk-2} yields that
\begin{align}
\label{d_t7}	\beta( \alpha_{11}\lambda_1+\alpha_{12}\lambda_2) - \lambda_1 + \alpha_{11}s + \alpha_{12}\cdot 2s &= 0\\
\label{d_t8}	\beta( \alpha_{21}\lambda_1+\alpha_{22}\lambda_2) - \lambda_2 + \alpha_{21}\cdot0 + \alpha_{22}\cdot s&= 0 
\end{align}

Suppose $\lambda_1 = \lambda_2 = \omega$, then $(1+\alpha_{12})s = (1-\beta)\omega = \alpha_{22}s $. But $s>0$ and $1+\alpha_{12}>1>\alpha_{22}$, thus  $(1+\alpha_{12})s < \alpha_{22}s $. Therefore we cannot have $ \delta^{alt*}_1> 0$ and $ \delta^{alt*}_2 > 0$.  Suppose $\lambda_2 = \omega$. Then solving the system of equations gives $\lambda_1 = \frac{1+\alpha_{12}-\beta\alpha_{22}}{\beta(2\alpha_{21}+\alpha_{12}-1)+\alpha_{22}}\cdot \omega >\omega$, which contradicts the KKT condition \eqref{eq:dk-1}. Hence $\lambda_2 < \omega$ implies that $\delta^{alt*}_2 = 0$ and thus $ \delta^{alt*}_1>0$. Therefore, $\lambda_1 = \omega$, $\lambda_2 =  \frac{\beta(2\alpha_{21}+\alpha_{12}-1)+\alpha_{22}}{1+\alpha_{12}-\beta\alpha_{22}}\cdot \omega$ and $s = (1-\beta)-\frac{(1-\beta)^2\alpha_{12}}{1+\alpha_{12}-\beta\alpha_{22}}\cdot \omega$.

Let $p^{ori}_i = p^{alt*}_i$, $y^{ori}_{ij} = 0$ for all $i,j$; $ x^{ori}_1 = x^{ori}_2 = \delta^{ori}_1 = \delta^{ori}_2 =0$, $z^{ori}_1 = d^{alt*}_1$ and $z^{ori}_2 = \alpha_{12}d^{alt*}_1+\alpha_{22}d^{alt*}_2$; $r^{ori}_{21} = \alpha_{12}d^{alt*}_1-\alpha_{21}d^{alt*}_2$ and $r^{ori}_{11} = r^{ori}_{12} = r^{ori}_{22}=0$ (notice that $r^{ori}_{21} >0$ since $\alpha_{12}\Delta_1> -\alpha_{22}\Delta_2$ implies that $\alpha_{12}d^{alt*}_1+\alpha_{22}d^{alt*}_2 > \alpha_{12}x^{alt*}_1+\alpha_{22}x^{alt*}_2 = \frac{x^{alt*}_2}{\beta}>x^{alt*}_2>d^{alt*}_2 $ and thus $\alpha_{12}d^{alt*}_1-\alpha_{21}d^{alt*}_2 >0$). Then $ \mathbf{u}^{ori} = \left\{p^{ori}_i, \delta^{ori}_i, z^{ori}_i, x^{ori}_i, y^{ori}_{ij}, r^{ori}_{ij}\right\}_{i,j=1}^2$ would be a feasible solution for \eqref{eq:opt_d1}. This solution decreases the cost by $\omega \cdot (\delta^{alt*}_1- \delta^{ori}_1+\delta^{alt*}_2 -\delta^{ori}_2) = (1-\beta)\omega (x^{alt*}_1+x^{alt*}_2)$, increases the cost by $s\cdot(z^{ori}_1- z^{alt*}_1+z^{ori}_2 -z^{alt*}_2) = s\cdot (x^{alt*}_1+ \alpha_{12}x^{alt*}_1+\alpha_{22}x^{alt*}_2) =  (1-\beta)\omega (x^{alt*}_1+x^{alt*}_2)$. The net profit is not changing, hence there always exists a solution for the original optimization problem that has the same profit which proves the claim.
\end{proof}
\end{proofcomment}

Turning our attention to AV priority assignment case, consider the optimization problem
\begin{align}
\nonumber \max_{\{p_i,\delta_i,x_i,y_{ij},z_i,r_{ij}\}}&\sum_{i=1}^n p_i \theta_i (1-F(p_i)) - \omega\sum_{i=1}^n \delta_i - s\sum_{i=1}^n z_i\\
\nonumber \text{s.t.}\quad d_i=&\theta_i (1-F(p_i))  \\
\nonumber  x_i=&\beta \left[\sum_{j} \alpha_{ji} (d_j-z_j) + \sum_{j} y_{ji} \right] + \delta_i\\
\nonumber \sum_{j=1}^n y_{ij}=& x_i -( d_i-z_i) \\
\nonumber  z_i=&\sum_{j=1}^n \alpha_{ji} z_j\\
  \label{eq:opt_a2} p_i,&\delta_i,z_i,x_i,y_{ij}\geq 0 \qquad \forall i,j.
\end{align}
Similar to above, we regard \eqref{eq:opt_a1} as the \emph{original} optimization problem and \eqref{eq:opt_a2} as the \emph{alternative} optimization problem for AV priority assignment. The next theorem mirrors Theorem \ref{thm:equiv_driver}.

\begin{thm}
\label{thm:equiv_AV}
Consider the original optimization problem \eqref{eq:opt_a1} and alternative optimization problem \eqref{eq:opt_a2}. Let
\begin{align}
  \label{eq:a-ori-sol}
  \mathbf{u}^{ori*} = \left\{p_i^{ori*}, \delta_i^{ori*}, z_i^{ori*}, x_i^{ori*}, y_{ij}^{ori*}, r_{ij}^{ori*}\right\}_{i,j=1}^n
\end{align}
be an optimal solution for \eqref{eq:opt_a1} and 
\begin{align}
  \label{eq:a-alt-sol}
  \mathbf{u}^{alt*} = \left\{p_i^{alt*}, \delta_i^{alt*}, z_i^{alt*}, x_i^{alt*}, y_{ij}^{alt*}\right\}_{i,j=1}^n
\end{align}
be an optimal solution for \eqref{eq:opt_a2}. 
Then the following holds under Assumptions \ref{assum:all} and \ref{assum:obj_concave}:
\begin{itemize}
\item The original optimization problem and the alternative problem obtain the same optimal profits for all possible choices of $\beta$, \change{$s$, $\omega$} and adjacency matrix $\mathbf{A}$.

\item The optimal solutions satisfy $x^{ori*} =  x^{alt*}$, $z^{ori*} = z^{alt*}$, $p^{ori*} = p^{alt*}$ and $\delta^{ori*} = \delta^{alt*}$.

\item If $\theta_i (1-F(p^{ori*}_i))\leq z_i^{ori*}$ for all $i$ in the original optimization problem, then $x_i^{ori*}=0$ for all $i$ and setting $y^{alt*}_{ij} = r^{ori*}_{ij}$ for all $i,j$ constitutes an optimal solution for the alternative problem.

\item If $ \theta_i (1-F(p^{alt*}_i))\leq z_i^{alt*}$ for all $i$ in the alternative optimization problem, then $x_i^{alt*}=0$ for all $i$ and setting $r^{ori*}_{ij} = y^{alt*}_{ij}$, $y^{ori*}_{ij} = 0$ constitutes an optimal solution for the original optimization problem. 
\end{itemize}
\end{thm}

\begin{proofcomment}
\begin{proof}

The proving strategy is the same as Theorem \ref{thm:equiv_driver}. 
Let $\profit^{ori*}$ and $\profit^{alt*}$ represent the optimal profits of the two problems \eqref{eq:opt_a1} and \eqref{eq:opt_a2}, respectively, and let $d^{ori*}_i=\theta_i (1-F(p^{ori*}_i))$ and $d^{alt*}_i= \theta_i (1-F(p^{alt*}_i))$.

The  KKT conditions related to all of the decision variables (except for the variable $p_i$ since $F(p_i)$ can be some general function of $p_i$) are:
\begin{align}
  \label{eq:ak-1} \text{(constraints on $\delta_i$)}:&-\omega +\lambda_i \leq 0\\ 
  \label{eq:ak-2} \text{(constraints on $x_i$)}: &-\lambda_i+\gamma_i\leq 0 \\
  \label{eq:ak-3} \text{(constraints on $z_i$)}: &-s - \sum_j \alpha_{ij} (\beta \lambda_j-\mu_j) + \gamma_i-\mu_i\leq 0\\
  \label{eq:ak-4} \text{(constraints on $y_{ij}$)}: &\beta \lambda_j-\gamma_i \leq 0.
\end{align}
Notice that for any of the inequalities, the equality holds if the corresponding variable is greater than zero.

To prove that the optimal profits of the two problems are equal, we first show that $\profit^{ori*} \leq  \profit^{alt*}$ and then $\profit^{ori*} \geq  \profit^{alt*}$.
In both directions, the first two cases ($d_i \leq x_i$ for all $i$ and $d_i \geq x_i$ for all $i$) use exactly the same method as the proof in Theorem \ref{thm:equiv_driver}, hence we omit those details, and only consider the third case to prove $\profit^{ori*} \leq  \profit^{alt*}$.


\underline{Case 3:} There exists some location $i$ such that $z^{ori*}_i < d^{ori*}_i$ and some location $j$ such that $z^{ori*}_j > d^{ori*}_j$. We will prove that the optimal solution for the original optimization problem  \eqref{eq:opt_a1} will not fall in this case. 

Suppose there exist some location such that $z^{ori*}_i < d^{ori*}_i$,  and let $I_1=\{i: z^{ori*}_i < d^{ori*}_i \}$ and $I_2=\{i: z^{ori*}_i \geq d^{ori*}_i \}$. We will show that for all $i \in I_2$, $z^{ori*}_i = d^{ori*}_i$. 
 We can consider an aggregated network with locations $1$ and $2$ representing the combined locations in $I_1$ and $I_2$, respectively. Knowing $z_1 < d_1$ and $z_2 \geq d_2$, then for any $d_2$, $z_2 = d_2$ will constitute a feasible solution for \eqref{eq:opt_a1}. Moreover, any $z_2$ such that $z_2 > d_2$ will increase the cost and thus decrease the profit for \eqref{eq:opt_a1}. Hence $z_2 = d_2$ is optimal. Therefore case 3 will not constitute an optimal solution for \eqref{eq:opt_a1}.

Next we consider the third case for proving $\profit^{ori*} \geq  \profit^{alt*}$.

\underline{Case 3:} There exists some location $i$ such that $z^{alt*}_i < d^{alt*}_i$ and some location $j$ such that $z^{alt*}_j > d^{alt*}_j$. We will prove that the optimal solution for the alternative optimization problem will not fall in this case. 

Suppose there exists some location such that $z^{alt*}_i < d^{alt*}_i$, and let $I_1=\{i: z^{alt*}_i < d^{alt*}_i \}$ and $I_2=\{i: z^{alt*}_i \geq d^{alt*}_i \}$. We will show that for all $i \in I_2$, $z^{ori*}_i = d^{ori*}_i$. 

As above, we can consider an aggregated network with locations $1$ and $2$ representing the combined locations in $I_1$ and $I_2$, respectively. We know that $z^{alt*}_1 < d^{alt*}_1$ and denote $\Delta_1 = d_1^{alt*}-z_1^{alt*}>0$. Moreover, suppose that $z^{alt*}_2 > d^{alt*}_2$ and $\Delta_2 = d_2^{alt*}-z_2^{alt*}<0$. We then rewrite the constraints in \eqref{eq:a_t1} as below:
\begin{align}
\nonumber x^{alt*}_1 = &\beta[\alpha_{11}\Delta_1 +\alpha_{21}\Delta_2 + (y^{alt*}_{11} +y^{alt*}_{21})  ] +\delta^{alt*}_1\\
\nonumber x^{alt*}_2 = &\beta[\alpha_{12}\Delta_1 +\alpha_{22}\Delta_2 + (y^{alt*}_{12} +y^{alt*}_{22})  ] + \delta^{alt*}_2 \\
\nonumber y^{alt*}_{11} +y^{alt*}_{12}=& x^{alt*}_1-\Delta_1 \\
\nonumber y^{alt*}_{21} +y^{alt*}_{22}=& x^{alt*}_2-\Delta_2 \\
\nonumber  z^{alt*}_1=&\alpha_{11}z^{alt*}_1+\alpha_{21}z^{alt*}_2\\
\nonumber  z^{alt*}_2=&\alpha_{12}z^{alt*}_1+\alpha_{22}z^{alt*}_2\\
  \label{eq:a_t1} p_i,&\delta_i,z_i,x_i,y_{ij}\geq 0 \qquad \forall i,j.
\end{align}

First notice that $x_1>0$ and $y_{21} +y_{22} >0$ since $\Delta_1>0$ and $\Delta_2<0$; then $x_1+x_2>0$ and thus $\delta_1+\delta_2 = (1-\beta)(x_1+x_2)>0$. Moreover, we will show below that $\delta_1+y_{21}>0$.
	
	Suppose that $\delta_1 = y_{21} = 0$. Since $y_{12} \geq 0$, then $y_{11} \leq x_1-\Delta_1$. Then, from \eqref{eq:a_t1}, $x_1 = \beta [\alpha_{11}\Delta_1 +\alpha_{21}\Delta_2 + y_{11}]\leq \beta [\alpha_{11}\Delta_1 +\alpha_{21}\Delta_2 + x_1-\Delta_1 ] = \beta [-\alpha_{12}\Delta_1 +\alpha_{21}\Delta_2 + x_1] < \beta x_1 < x_1$. This is a contradiction and thus  $\delta_1 + y_{21} > 0$.

We next show that when $z^{alt*}_1 < d^{alt*}_1$ and $z^{alt*}_2 > d^{alt*}_2$, we are always able to obtain a solution in the original optimization problem that achieves greater profit. Since we have already proved that $\phi^{ori*}\leq\phi^{alt*}$, then the solution that falls in this case will not be an optimal solution for the alternative optimization problem.   

Suppose $s > (1-\beta)\omega$.  We are able to obtain a higher profit by increasing the mass of HVs and decreasing the mass of AVs. In particular, this transformation to case 1 is accomplished by setting $d^{ori}_i = d^{alt}_i$, $r^{ori}_{ij} = 0$, and $y^{ori}_{ij} =y^{alt}_{ij}$ for all $i, j$;  $z_2^{ori} = d_2^{alt}$, $z^{ori}_1 = \frac{\alpha_{21}}{\alpha_{12}}z^{ori}_2 $, $x^{ori}_1 = x^{alt}_1-  \frac{\alpha_{21}}{\alpha_{12}} \Delta_2$, $x^{ori}_2 = x^{alt}_2- \Delta_2$, $\delta^{ori}_1 = \delta^{alt}_1 - (1-\beta) \frac{\alpha_{21}}{\alpha_{12}}\Delta_2$ and $\delta^{ori}_2 = \delta^{alt}_2 - (1-\beta)\Delta_2$. Then, it is straightforward to verify that $\mathbf{u}^{ori} = \left\{p_i^{ori}, \delta_i^{ori}, z_i^{ori}, x_i^{ori}, y_{ij}^{ori}, r_{ij}^{ori}\right\}_{i,j=1}^2$ satisfies all the constraints of \eqref{eq:opt_a1}, and hence it is a feasible solution for \eqref{eq:opt_a1}.

This modified solution keeps the demand $d_i$ and thus $p_i$ unchanged, decreases the cost incurred by AVs by $s\cdot(z^{alt*}_1- z^{ori}_1+z^{alt*}_2 -z^{ori}_2) = s\cdot(-\frac{\alpha_{21}}{\alpha_{12}}\Delta_2 - \Delta_2) = -s\cdot(1+\frac{\alpha_{21}}{\alpha_{12}})\Delta_2 < \omega(1-\beta)(1+\frac{\alpha_{21}}{\alpha_{12}})\Delta_2$, and increases the cost incurred by HVs by $\omega \cdot (\delta^{ori}_1- \delta^{alt}_1+\delta^{ori}_2 -\delta^{alt}_2) = \omega(1-\beta)(1+\frac{\alpha_{21}}{\alpha_{12}})\Delta_2$. The net profit increases, hence there always exists a solution for the original optimization problem that achieves a higher profit. Thus, the original solution is not optimal.

Now consider when $s \leq (1-\beta)\omega$. Suppose $x^{alt*}_2=0$. Then $y^{alt*}_{21}+y^{alt*}_{22} = -\Delta_2$; since $x^{alt*}_1>0$ (and thus $\lambda_1 = \gamma_1$), it must hold that $y^{alt*}_{11}=0$ by KKT conditions. Moreover, we show that $y^{alt*}_{12} =0$. Suppose $y^{alt*}_{12}>0$ so that $\beta \lambda_2-\gamma_1 = 0$. While $\gamma_1 = \lambda_1 \in [\beta\omega, \omega]$ (this is true if there exist $x^{alt*}_i>0$ for any $i$), we must have $\lambda_2 = \omega$ and $\gamma_1 = \lambda_1 = \beta\omega$. If $\delta^{alt*}_1=0$, then $y^{alt*}_{21}>0$, and thus $\beta \lambda_1-\gamma_2 = 0$. Hence $\gamma_2 = \beta^2\omega$. However, we require $\beta \lambda_2-\gamma_2 \leq 0$ while $\beta \lambda_2-\gamma_2 = \beta\omega-\beta^2\omega>0$. Therefore $\delta^*_1>0$. But then we obtain $\lambda_1 = \omega$ by KKT conditions, which contradicts with the fact that $\lambda_1 = \beta\omega$. Therefore, $y^{alt*}_{12}=0$.

Since $y^{alt*}_{11} = y^{alt*}_{12}=0$, it holds that $x^{alt*}_1 = \Delta_1$. Since $x^{alt*}_2 = 0$, we can thus compute $y^{alt*}_{22} = -\alpha_{12}\Delta_1-\alpha_{22}\Delta_2-\frac{\delta^{alt*}_2}{\beta}\geq 0$, $y^{alt*}_{21} = \alpha_{12}\Delta_1-\alpha_{21}\Delta_2+\frac{\delta^{alt*}_2}{\beta}>0$. Notice that  $y^{alt*}_{21} >0$ because $\Delta_1>0$ and $\Delta_2<0$. Also, $\delta^{alt*}_1+\delta^{alt*}_2 = (1-\beta)(x^{alt*}_1+x^{alt*}_2) =(1-\beta)\Delta_1$.

Now consider the solution for the original optimization problem by setting $d^{ori}_i = d^{alt*}_i$, $x^{ori}_i = \delta^{ori}_i = y^{ori}_{ij} = 0$ for all $i,j$. Then a feasible solution of \eqref{eq:opt_a1} is obtained according to $z^{ori}_1 = d^{alt*}_1$, $z^{ori}_2 = z^{alt*}_2$, $r^{ori}_{11} = r^{ori}_{12}  = 0$, $r^{ori}_{21} = \alpha_{12}\Delta_1-\alpha_{21}\Delta_2$ and $r^{ori}_{22} = -\alpha_{12}\Delta_1-\alpha_{22}\Delta_2= y^{alt*}_{22}+\frac{\delta^{alt*}_2}{\beta}>0$. Then $\mathbf{u}^{ori} = \left\{p_i^{ori}, \delta_i^{ori}, z_i^{ori}, x_i^{ori}, y_{ij}^{ori}, r_{ij}^{ori}\right\}_{i,j=1}^2$.

Considering the cost of this modified solution compared to the original solution, the cost increases by $s\cdot(z^{ori}_1+z^{ori}_2)-s\cdot(z^{alt*}_1+z^{alt*}_2) = s\cdot \Delta_1$ and subsequently decreases by $\omega(\delta^{alt*}_1+\delta^{alt*}_2)-\omega(\delta^{ori}_1+\delta^{ori}_2) = (1-\beta)\omega \Delta_1 > s\cdot\Delta_1$. Since we have already proved that $\profit^{ori*} \leq  \profit^{alt*}$, this implies the original solution is not optimal, a contradiction.

Therefore, $x^{alt*}_2>0$, and by KKT conditions, $y^{alt}_{22} = y^{alt}_{11} = y^{alt*}_{12} = \delta^{alt*}_2 = 0$ and $y^{alt*}_{21} > 0$. Moreover, $\gamma_1 = \lambda_1 =  \omega$ and  $\gamma_2 = \lambda_2 = \beta \omega$. Hence $x^{alt*}_2 = \beta(\alpha_{12}\Delta_1+\alpha_{22}\Delta_2)>0$ and $x^{alt*}_1 = \Delta_1 $.

By \eqref{eq:ak-3}, we  have
\begin{align}
-s-\beta(\alpha_{11}\lambda_1+\alpha_{12}\lambda_2)+\gamma_1+(\alpha_{11}\mu_1+\alpha_{12}\mu_2)-\mu_1 &= 0\\
-s-\beta(\alpha_{21}\lambda_1+\alpha_{22}\lambda_2)+\gamma_2+(\alpha_{21}\mu_1+\alpha_{22}\mu_2)-\mu_2 &= 0.
\end{align}

Hence $-s+(1-\alpha_{11}\beta-\alpha_{12}\beta^2)\omega+\alpha_{12}(\mu_2-\mu_1) = 0$ and $-s+\alpha_{22}(1-\beta)\beta\omega+\alpha_{21}(\mu_1-\mu_2) = 0$. By adding coefficients $\alpha_{21}$ and $\alpha_{12}$, we obtain $-(\alpha_{12}+\alpha_{21})s+(1-\alpha_{11}\beta-\alpha_{12}\beta^2) \alpha_{21} \omega+ \alpha_{12}\alpha_{22}(1-\beta)\beta\omega=0$. By simplification, we then have $s = \frac{ (1+\beta)(\alpha_{21}-\alpha_{12}\beta)}{\alpha_{12}+\alpha_{21}}\cdot \omega$.

At the same time,  the equation $x_i=\beta \left[\sum_{j} \alpha_{ji} (d_j-z_j) + \sum_{j} y_{ji} \right] + \delta_i$ can be reformulated into $x_i=\beta \left[\sum_{j} \alpha_{ji} d_j-z_i + \sum_{j} y_{ji} \right] + \delta_i$, and hence the KKT condition corresponding to the reformulated optimization problem becomes
\begin{align}
  \label{eq:ak-5} \text{(constraints on $\delta_i$)}:&-\omega +\lambda^1_i \leq 0\\ 
  \label{eq:ak-6} \text{(constraints on $x_i$)}: &-\lambda^1_i+\gamma^1_i\leq 0 \\
  \label{eq:ak-7} \text{(constraints on $z_i$)}: &-s + \sum_j \alpha_{ij} \mu^1_j - \beta \lambda^1_i + \gamma^1_i-\mu^1_i\leq 0\\
  \label{eq:ak-8} \text{(constraints on $y_{ij}$)}: &\beta \lambda^1_j-\gamma^1_i \leq 0.
\end{align}

By the same process as before, we obtain $\gamma^1_1 = \lambda^1_1 =  \omega$, $\gamma^1_2 = \lambda^1_2 = \beta \omega$, and 
\begin{align}
  -s+(1-\beta)\lambda^1_1+(\alpha_{11}\mu^1_1+\alpha_{12}\mu^1_2)-\mu^1_1 &= 0 \\
  -s+(1-\beta)\lambda^1_2+(\alpha_{21}\mu^1_1+\alpha_{22}\mu^1_2)-\mu^1_2 &= 0.
\end{align}
Therefore, $s = \frac{(1-\beta)(\alpha_{21}+\alpha_{12}\beta)}{\alpha_{12}+\alpha_{21}}\cdot\omega$.

By establishing the equality  $s = \frac{(1-\beta)(\alpha_{21}+\alpha_{12}\beta)}{\alpha_{12}+\alpha_{21}}\cdot\omega = \frac{ (1+\beta)(\alpha_{21}-\alpha_{12}\beta)}{\alpha_{12}+\alpha_{21}}\cdot \omega$, we require $\alpha_{21} = \alpha_{12}$ and thus $s = \frac{\beta(1-\beta)\omega}{2}$.

Similar to the situation when $x^{alt*}_2 = 0$, we obtain a feasible solution $\mathbf{u}^{ori} = \left\{p_i^{ori}, \delta_i^{ori}, z_i^{ori}, x_i^{ori}, y_{ij}^{ori}, r_{ij}^{ori}\right\}_{i,j=1}^2$ for the original optimization problem by setting $d^{ori}_i = d^{alt*}_i$, $x^{ori}_i = \delta^{ori}_i = y^{ori}_{ij} = 0$ for all $i,j$; $z^{ori}_1 = d^{alt*}_1$, $z^{ori}_2 = \alpha_{12}d^{alt*}_1+\alpha_{22}d^{alt*}_2$, $r^{ori}_{11} = r^{ori}_{12}  = 0$, $r^{ori}_{21} = \alpha_{12}\Delta_1-\alpha_{21}\Delta_2$ and $r^{ori}_{22} = 0$. All constraints of \eqref{eq:opt_a1} are satisfied. 

The cost incurred by HVs is decreased by $\omega(\delta^{alt*}_1+\delta^{alt*}_2)-\omega(\delta^{ori}_1+\delta^{ori}_2) = (1-\beta)\omega (x^{alt*}_1+x^{alt*}_2) = \omega (1-\beta)(\Delta_1+ \beta(\alpha_{12}\Delta_1+\alpha_{22}\Delta_2))$ and the cost incurred by AVs  is increased by $s \cdot (z^{ori}_1+z^{ori}_2-z^{alt*}_1-z^{alt*}_2) = s \cdot (\Delta_1+\alpha_{12}\Delta_1+\alpha_{22}\Delta_2) =  \frac{\beta(1-\beta)\omega}{2}(\Delta_1+\alpha_{12}\Delta_1+\alpha_{22}\Delta_2) = \frac{\omega(1-\beta)}{2}(\beta\Delta_1+\beta(\alpha_{12}\Delta_1+\alpha_{22}\Delta_2)) <\omega (1-\beta)(\Delta_1+ \beta(\alpha_{12}\Delta_1+\alpha_{22}\Delta_2))$. Hence the cost decreases and the profit is not optimal for the original solution, a contradicition.

Therefore the optimal solution does not fall in case 3.
\end{proof}
\end{proofcomment}

Corollary \ref{cor:profit} follows from Theorems \ref{thm:equiv_driver} and \ref{thm:equiv_AV}.

\begin{cor}
\label{cor:profit}
  Under Assumptions \ref{assum:all} and \ref{assum:obj_concave}, the optimal profit for the mixed autonomy deployment under HV (resp., AV) priority assignment is no less than the optimal profit computed from \eqref{eq:opt_d1}/\eqref{eq:opt_a1} with the additional forced HV-only deployment constraint, \emph{i.e.}, the constraint $z_i=0$ for all $i$.
\end{cor}

\begin{proofcomment}
\begin{proof}
  The mixed autonomy optimization problem can be transformed into \eqref{eq:d2_1} by setting $\mathbf{z=0}$ and $\mathbf{r=0}$. Furthermore, \eqref{eq:d2_1} is exactly the optimization problem for the system without any AVs. Therefore, by letting $\mathbf{z=0}$ and $\mathbf{r=0}$ and the other variables equal to the optimal solution for the optimization problem for the system without AV, we obtain a feasible solution for the mixed autonomy system. Therefore the optimal profit for the mixed autonomy system will be no less than that of the system without autonomous system.
\end{proof}
\end{proofcomment}

Corollary \ref{cor:profit} emphasizes that in our model, the AVs will be introduced into the platform only if they increase the optimal profit for the platform.

\begin{proofsketchcomment}

\end{proofsketchcomment}

\section{The Relation between HV Priority and AV Priority Assignments}
\label{sec:D-A-model-relation}
Now that we have introduced the alternative optimization problems for maximizing the profits in both HV and AV priority assignments, we next compare the optimal profits for the two priority assignments. \change{The main result of this section is Theorem \ref{thm: driver = AV} which shows that the two priority assignments actually lead to the same optimal profits.}

\change{Before presenting the main theorem, we  first introduce some preliminary lemmas that are interesting in their own right.} In the remainder of the paper, we denote an optimal solution with superscript $*$, \emph{e.g.}, $x_i^*$.

The next lemma  establishes that \change{under HV priority assignment, if some location has departing AVs without passengers, then that location also does not have incoming AVs without passengers.}

\begin{lemma}
\label{lemma: driver I/O}
Consider the alternative optimization problem \eqref{eq:opt_d2} for HV priority assignment under Assumptions \ref{assum:all} and \ref{assum:obj_concave}. Suppose there exist some location $i$ such that both $x^{*}_i >0$ and $z^{*}_i >0$. Then $d^*_i \geq x^*_i$ for all $i$. Moreover, for any $i_0$, if there exists some location $j$ such that $r^*_{i_0 j}>0$, then $r^*_{j i_0} = 0$ for all $j$.
\end{lemma}

\begin{proofcomment}
\begin{proof}\

\underline{Step 1}: We first show that  $d^*_i \geq x^*_i$ for all $i$. This part follows similar to the corresponding part in Lemma \ref{lemma: AV I/O} which will be proved later.

\underline{Step 2}: We complete the proof by contradiction. Assume $i_0, j_0$ are locations that $r^*_{{i_0}{j_0}}>0$. By \eqref{eq:dk-4} we'll have $\mu_{j_0}-\gamma_{i_0}=0$. 

Since $\sum_{j=1}^n r_{ij}=z_i -(d_i-x_i )$ by \eqref{eq:opt_d2}, then $z_{i_0} = d_{i_0} - x_{i_0}+\sum_{j=1}^n r_{{i_0}j} = d_{i_0} - x_{i_0}+\sum_{j=1, j\neq{j_0}}^n r_{{i_0}j}+r_{{i_0}{j_0}}$. Since $r_{{i_0}j}\geq 0$ for all $j$, $r^*_{{i_0}{j_0}}>0$ and from step 1 we have $d^*_{i_0} \geq x^*_{i_0}$, then $z^*_{i_0}>0$. And \eqref{eq:dk-3} gives that $ \gamma_{i_0}-\mu_{i_0} = s$. Combining the two results yields that $\mu_{j_0}-\mu_{i_0} =s$.

Suppose there exists a location $j$ that $r^*_{j{i_0}}>0$, then $\mu_{i_0}-\gamma_{j}=0$. Hence $\mu_{i_0}=\gamma_{j}$ and $\mu_{j_0}-\gamma_{j}=\mu_{j_0}-\mu_{i_0} =s>0$ which contradicts \eqref{eq:dk-4}. Therefore, for any location $j$, $r^*_{j{i_0}}=0$.
\end{proof}
\end{proofcomment}

\change{Next, we show that }if it is optimal for the platform to use both HVs and AVs at some location, then every vehicle in the network will be assigned to a ride.

\begin{lemma}
\label{prop: driver mix}
For optimization problem \eqref{eq:opt_d2} under Assumption \ref{assum:all} and \ref{assum:obj_concave}, if there exists a location $i_0$ such that $x^*_{i_0}>0$ and $z^*_{i_0}>0$, then $r^*_{ij} = 0$ for all $i,j$.
\end{lemma}

\begin{proofcomment}
\begin{proof}\

Since there exists a location $i$ such that $x^*_{i}>0$ and $z^*_{i}>0$, from Lemma \ref{lemma: driver I/O} we know that $d^*_i \geq x^*_i$ for all $i$.

Suppose there exist a location $i_0$ such that $r^*_{{i_0}j}>0$. First, we partition the $n$ locations into two groups: $I_1 = \left\{ i: i\neq i_0\right\}$, $I_2 = \left\{ i_0\right\}$. Then we  aggregate those into a 2-location system with locations $1$ and $2$ such that $\alpha_{22} = 0$, $\alpha_{21} = 1$. 

\noindent \underline{Step 1}: We show $r^*_{11} =r^*_{12} = r^*_{22} =0$, $r^*_{21}>0$.

Notice that since $r^*_{{i_0}j}>0$, then $r^*_{21}>0$. Hence by Lemma \ref{lemma: driver I/O}, $r^*_{12} = r^*_{22} =0$. Moreover, since $z_1 = \alpha_{11}(d_1-x_1) +\alpha_{21}(d_2-x_2) +r_{11}+r_{21}$ and $d^*_i\geq x^*_i$ for $i=1,2$, then $z^*_1>0$ and $\gamma_1-\mu_1 = s$ by \eqref{eq:dk-3}. From \eqref{eq:dk-4}, $\mu_1 - \gamma_1 = -s<0$ implies that $r^*_{11} = 0$.

\noindent \underline{Step 2}: We show that $\delta^*_2 =0$ and $\delta^*_1 >0$ using KKT conditions.

To reason about the 2-group problem, first rewrite the optimization constraints below by combining with the conditions $\alpha_{22} = 0$, $\alpha_{21} = 1$.

\begin{align}
\nonumber  x_1&=\beta (\alpha_{11}x_1 + x_2) + \delta_1\\
\nonumber  x_2&=\beta \alpha_{12}x_1 + \delta_2\\
\nonumber  z_1&=\alpha_{11} (d_1-x_1) + (d_2-x_2)+ r_{11}+r_{21}\\
\nonumber  z_2&=\alpha_{12} (d_1-x_1)+ r_{12}+r_{22}\\
\nonumber  r_{11}+r_{12}&=z_1 + x_1 - d_1\\
\nonumber  r_{21}+r_{22}&=z_2 + x_2 - d_2\\
 \label{eq:18} \delta_i,x_i,z_i,r_{ij}&\geq 0 \qquad \forall i,j.
\end{align}

Clearly, as $x^*_i>0$ for $i = 1 $ or $2$, then $x^*_1>0$ and $x^*_2>0$ since $\alpha_{12}>0$ when the actual ride-sharing network has no less then two locations and is strongly connected. Similarly, since there exists a location $i$ such that $x^*_{i}>0$ and $z^*_{i}>0$, then $d^*_i-x^*_i>0$ and $d^*_1-x^*_1>0$ or $d^*_2-x^*_2>0$. Hence $z^*_1>0$  and $z_2 = r_{21}+r_{22} + d_2 - x_2$ implies that $z^*_2>0$. 

We can therefore conclude the corresponding KKT conditions:
\begin{align*}
\nonumber   r_{21}>0 &\Rightarrow \mu_1-\gamma_2=0 \\
\nonumber  z_{1}>0 &\Rightarrow \gamma_1-\mu_1=s \\
\nonumber  z_{2}>0 &\Rightarrow \gamma_2-\mu_2=s \\
\nonumber  x_{1}>0 &\Rightarrow \alpha_{11}(\beta\lambda_1-\mu_1)+\alpha_{12}(\beta\lambda_2-\mu_2)-\lambda_1+\gamma_1=0 \\
\nonumber  x_{2}>0 &\Rightarrow (\beta\lambda_1-\mu_1)-\lambda_2+\gamma_2=0.
\end{align*}

Notice that  the first 3 equations above imply that $\gamma_1-\mu_2 = 2s +\mu_1 - \gamma_2 = 2s$. 
By recombination of the equations, we  derive
\begin{align}
\nonumber	\beta(\alpha_{11}\lambda_1 +\alpha_{12}\lambda_2)-\lambda_1+\alpha_{11}(\gamma_1-\mu_1)+\alpha_{12}(\gamma_1-\mu_2)&=0\\
\nonumber	\beta(\alpha_{11}\lambda_1 +\alpha_{12}\lambda_2)-\lambda_1+\alpha_{11}\cdot s+\alpha_{12}\cdot 2s&=0\\
\label{eq:18-1}	\beta(\alpha_{11}\lambda_1 +\alpha_{12}\lambda_2)-\lambda_1+(1+\alpha_{12})s&=0
\end{align}
and 
\begin{align}
\nonumber	\beta\lambda_1 -\lambda_2 - (\mu_1-\gamma_2) &=0\\
\label{eq:18-2}	\beta\lambda_1 -\lambda_2 &=0.
\end{align}

Since $\delta_1+\delta_2 = (1-\beta)(x_1+x_2)$ and now $x^*_1+x^*_2>0$, then $\delta^*_1+\delta^*_2>0$.
Suppose $\delta^*_2>0$, then by $\eqref{eq:dk-1}$, $\lambda_2 =\omega$, and hence $\lambda_1 = \frac{\lambda_2}{\beta}= \frac{\omega}{\beta}>\omega$, which contradicts the KKT condition. Hence $\delta^*_2 =0$ and thus $\delta^*_1 >0$.

\noindent \underline{Step 3}: Determine the range of $s$ that satisfies the given conditions.

Since $\delta^*_1 >0$ then $\lambda_1 =\omega$ and thus $\lambda_2 = \beta\lambda_1 = \beta \omega$. Substituting those into \eqref{eq:18-1} yields that
\begin{align}
 s &= -\frac{\alpha_{11}\beta\omega+\alpha_{12}\beta^2\omega-\omega}{1+\alpha_{12}} \\
 &= \frac{(1-\beta)(1+\alpha_{12}\beta)}{1+\alpha_{12}}\cdot \omega   .
\end{align}

Therefore, $s = \frac{(1-\beta)(1+\alpha_{12}\beta)}{1+\alpha_{12}}\cdot \omega$ is the only value that is feasible.

\noindent \underline{Step 4}: We show that it is possible for the platform to realize the same profit using only AVs ($x_i=0, z_i>0$ for all $i$).
Now that
\begin{align}
\nonumber  x^*_1&=\beta (\alpha_{11}x_1 + x_2) + \delta^*_1\\
\nonumber  x^*_2&=\beta \alpha_{12}x^*_1 \\
\nonumber  z^*_1&=\alpha_{11} (d^*_1-x^*_1) + (d^*_2-x^*_2)+ r^*_{21}\\
\nonumber  z^*_2&=\alpha_{12} (d^*_1-x^*_1)\\
\nonumber  0&=z^*_1 + x^*_1 - d^*_1\\
\nonumber  r^*_{21} &=z^*_2 + x^*_2 - d^*_2,
\end{align}
suppose $d^*_1 \leq d^*_2$. Since $x^*_2 = \beta\alpha_{12}x^*_1 < x^*_1$ and $z^*_1 = d^*_1-x^*_1 =\alpha_{11} (d^*_1-x^*_1) + (d^*_2-x^*_2)+ r^*_{21}$, then $d^*_1-x^*_1\geq d^*_2-x^*_2 =  d^*_2-\beta \alpha_{12}x^*_1 > d^*_2 - x^*_1$. This implies that $d^*_1 > d^*_2$, which contradicts the assumption. Therefore $d^*_1 > d^*_2$. 

Moreover, since $z^*_2 = r^*_{21} +  d^*_2 -  x^*_2  > d^*_2 -  x^*_2$, then $ z^*_2=\alpha_{12} (d^*_1-x^*_1)>d^*_2 -  x^*_2 \Rightarrow \alpha_{12} d^*_1 > d^*_2 -  x^*_2 + \alpha_{12} x^*_1  = d^*_2+(1-\beta)\alpha_{12} x^*_1$. We can also reformulate that $\delta^*_1 = (1-\beta)(x^*_1+x^*_2) = (1-\beta)(1+\beta\alpha_{12})x^*_1$ and $z^*_1+z^*_2 = (1+ \alpha_{12} )(d^*_1-x^*_1)$.

It is straightforward to verify that 
\begin{equation*}
    \left\{z_1 = d^*_1, z_2 = \alpha_{12} d^*_1, x_1 = x_2 = \delta_1 = \delta_2 =0, r_{ij} =  r^*_{ij}  \right\}
\end{equation*}
is also a feasible solution for the problem.

We now consider the modified costs under this alternative feasible solution. The increase of the cost is 
\begin{align}
\nonumber    &(z_1+z_2)\cdot s - (z^*_1+z^*_2)\cdot s \\
   \nonumber &= [d^*_1+\alpha_{12} d^*_1 - (1+ \alpha_{12} )(d^*_1-x^*_1)]\cdot s \\
   \nonumber &= (1+ \alpha_{12} )x^*_1\cdot  \frac{(1-\beta)(1+\alpha_{12}\beta)}{1+\alpha_{12}}\cdot \omega \\
    &= (1-\beta)(1+\alpha_{12}\beta)x^*_1\cdot \omega,
\end{align}
and the cost is subsequently decreased by $(\delta^*_1+\delta^*_2)\cdot \omega - (\delta_1+\delta_2)\cdot \omega = \delta^*_1\omega = (1-\beta)(1+\beta\alpha_{12})x^*_1\omega $. Thus the total cost does not change while the prices and demands are also unchanged. Therefore the profit is not changed.

Hence, it is possible to achieve the same profit using only AVs.

\noindent \underline{Step 5}: We next complete the proof by contradiction. Denote the solutions above as $\mathbf{u}^{d*}_{x>0,z>0}$ for the mixed case of both HVs and AVs and by $\mathbf{u}^{d*}_{x=0,z>0}$ for the case with only AVs. Denote the optimal profit obtained in these two scenarios as $\pi_m$ and $\pi_{AV}$, respectively, and from Step 4 we know $\pi_m = \pi_{AV}$. Consider the alternative form of AV priority assignment optimization problem \eqref{eq:opt_a2}. 

Suppose with the same $\omega, s, \beta$ and $\alpha_{ij}$ for all $i,j$, the optimal solution for AV priority assignment falls into the mixed autonomy case with $\mathbf{u}^{a*}_{x>0,z>0}$. Notice that since $y^{a*}_{ij, x>0,z>0} = 0$ for all $i,j$, then the solution $\mathbf{u}^{a*}_{x>0,z>0}$ is feasible for HV priority assignment by substituting $r^{d}_{ij} $ with $y^{a*}_{ij, x>0,z>0} $, and moreover the profit will be  exactly the same. Additionally, the solution $\mathbf{u}^{d*}_{x>0,z>0}$ is also feasible for AV priority assignment by substituting $y^{a}_{ij} $ with $r^{d*}_{ij, x>0,z>0}$ with the profit $\hat{\pi}_m$. However, since there exist $i,j$ such that $r^{d*}_{ij, x>0,z>0}>0$, and from Lemma \ref{prop: AV mix} (as we will prove later) we know that this is not optimal for AV priority assignment, it follows that $\pi_m < \hat{\pi}_m$. Hence $\pi_m$ is not an  optimal profit for HV priority assignment, which gives the contradiction.
	
Suppose the optimal solution $\mathbf{u}^{a*}_{x=0,z>0}$ for AV priority assignment falls into the pure-AV case, \emph{i.e.}, $x_i=0$ for all $i$. Again, $\mathbf{u}^{d*}_{x>0,z>0}$ is feasible for AV priority assignment. Moreover, under the case with only AVs, the two optimization problems are exactly the same by substituting $r^d_{ij}$ with $y^a_{ij}$. Therefore $\mathbf{u}^{d*}_{x=0,z>0}$ and $\mathbf{u}^{a*}_{x=0,z>0}$ yield the same profit, denoted as $\pi_{AV}$. However, since $\mathbf{u}^{d*}_{x>0,z>0}$ cannot be optimal for AV priority assignment as shown above, $\pi_m<\pi_{AV}$ which contradicts the above result that $\pi_m=\pi_{AV}$.

Finally, if the optimal solution $\mathbf{u}^{a*}_{x>0,z=0}$ for AV priority assignment falls into the pure-HV case, i.e., $z_i=0$ for all $i$, then the optimal profit gained from this solution, denoted as $\pi_{HV}$, will be greater than $\pi_m$ (since $\pi_m$ is not the optimal profit). Moreover, since the solution will also be feasible for HV priority optimization problem, then $\pi_{HV}$ is also attainable for HV priority assignment. This contradicts the result that $\pi_{m}$ is the optimal profit for HV priority assignment.

Therefore, $\mathbf{u}^{d*}_{x>0,z>0}$ cannot be the optimal solution for \eqref{eq:opt_d2} and our assumption that there exist $i,j$ such that $r^*_{ij}>0$ is false. Hence, in the situation under consideration, $r^*_{ij}=0$ for all $i,j$.
\end{proof}
\end{proofcomment}

Similar properties exist under AV priority assignment, as summarized in the following lemmas.

\begin{lemma}
\label{lemma: AV I/O}
Consider the alternative optimization problem \eqref{eq:opt_a2} for AV priority assignment under Assumptions \ref{assum:all} and \ref{assum:obj_concave}. 
Suppose there exist some location $i$ such that both $x^{*}_i >0$ and $z^{*}_i >0$. Then $d^*_i \geq z^*_i$ for all $i$. Moreover, for any $i_0$, if there exist some location $j$ such that $y_{i_0 j}>0$, then $y_{j i_0} = 0$ for all $j$.
\end{lemma}

\begin{proofcomment}
\begin{proof} \

\underline{Step 1}: We show that  $d^*_i \geq z^*_i$ for all $i$. Assume location $i_{>0}\in\{1,\ldots,n\}$ is such that $x^{*}_{i_{>0}} >0$ and $z^{*}_{i_{>0}}>0$. From the construction of the model, we know that the platform uses AVs only to meet the excess demand, hence $d^*_{i_{>0}} > z^*_{i_{>0}}$. Therefore, from Theorem \ref{thm:equiv_AV}, we know that for the optimal problem \eqref{eq:opt_a2}, $d^*_{i_{>0}} > z^*_{i_{>0}}$. Moreover, in the proof of the theorem, we have also shown that the mixed case where there exist some locations such that $d_i>z_i$ and some locations such that $d_i<z_i$ will not be the optimal solution. Thus, it follows that $d^*_i \geq z^*_i$ for all $i$ under this circumstance.

\underline{Step 2}: We complete the proof by contradiction. Assume $i_0, j_0$ are locations such that $y^*_{{i_0}{j_0}}>0$. By \eqref{eq:ak-4}, we have $\beta \lambda_{j_0}-\gamma_{i_0}=0$. Since $\sum_{j=1}^n y_{ij}=x_i -(d_i-z_i )$ by \eqref{eq:opt_a2}, then $x_{i_0} = d_{i_0} - z_{i_0}+\sum_{j=1}^n y_{{i_0}j} = d_{i_0} - z_{i_0}+\sum_{j=1, j\neq{j_0}}^n y_{{i_0}j}+y_{{i_0}{j_0}}$. Since $y_{{i_0}j}\geq 0$ for all $j$, $y^*_{{i_0}{j_0}}>0$, and from Step 1 above we have $d^*_{i_0} \geq z^*_{i_0}$, then $x^*_{i_0}>0$. Therefore, \eqref{eq:ak-2} gives that $ \gamma_{i_0}=\lambda_{i_0}$.

Notice also \eqref{eq:ak-1}, \eqref{eq:ak-2} and \eqref{eq:ak-4} together indicate that $\lambda_i \in [\beta\omega, \omega]$ and $\gamma_i \in [\beta\omega, \omega]$ for all $i$ when there exists at least one location $i^{'}$ such that $\delta_{i^{'}}>0$ (or $x_{i^{'}}>0$). Therefore, $\lambda_{j_0} = \omega,  \gamma_{i_0} = \beta\omega$ is the only possible choice. Thus $ \gamma_{i_0}=\lambda_{i_0} = \beta\omega$.

Suppose there exists a location $j$ such that $y^*_{j{i_0}}>0$. Then $\beta \lambda_{i_0}-\gamma_{j}=0$. This indicates that $\lambda_{i_0} = \omega$ and $  \gamma_{j} = \beta\omega$, which contradicts the result $\lambda_{i_0} = \beta\omega$ obtained above.
Therefore, for any location $j$, $y^*_{j{i_0}}=0$. 
\end{proof}
\end{proofcomment}

\begin{lemma}
\label{prop: AV mix}
For optimization problem \eqref{eq:opt_a2} under Assumptions \ref{assum:all} and \ref{assum:obj_concave}, if there exists a location $i_0$ such that $x^*_{i_0}>0$ and $z^*_{i_0}>0$, then $y^*_{ij} = 0$ for all $i,j$.
\end{lemma}

\begin{proofcomment}
\begin{proof}\

We  partition the locations into two groups: $I_1 = \left\{i: y^*_{ij} =0 \quad \forall j \right\}$ and  $I_2 = \left\{i: \exists j \quad y^*_{ij} >0 \right\}$. By aggregating these groups into two locations, we henceforth regard this as a two-location problem indexed by $1$ and $2$.
By Lemma \ref{lemma: AV I/O}, we know that $y^*_{22} = 0$, $y^*_{21}>0$ and $d^*_i \geq z^*_i$ for $i=1,2$.

As $z^*_1>0$ or $z^*_2>0$ and $z_i=\sum_{j=1}^2 \alpha_{ji} z_j$, since the network is strongly connected, then $z^*_1>0$ and $z^*_2>0$. Knowing $d^*_2 \geq z^*_2$, $x_2 = (d_2-z_2) + y_{21}+y_{22}$ and $y^*_{21}>0$ implies that $x^*_2>0$; similarly, $x^*_1 = \beta[\alpha_{11}(d^*_1-z^*_1)+\alpha_{21}(d^*_2-z^*_2)+y^*_{11}+y^*_{21}]>0$. Moreover, \eqref{eq:ak-2} implies that $\gamma_1=\lambda_1$ and $\gamma_2=\lambda_2$. 

Also, since $\delta_1+\delta_2 = (1-\beta)(x_1+x_2)$, then there exists $i\in\{1,2\}$ such that $\delta^*_i>0$ and hence $\lambda_i = \omega$ by \eqref{eq:ak-1}. Combining with \eqref{eq:ak-2} and \eqref{eq:ak-4}, we know that $\lambda_i \in [\beta\omega, \omega]$ and $\gamma_i \in [\beta\omega, \omega]$. Since $y^*_{21}>0$, then $\beta\lambda_1-\gamma_2=0$ which indicates that $\lambda_1=\omega = \gamma_1$ and $\gamma_2 = \beta\omega = \lambda_2$. We  further conclude that $\delta^*_2=0$ and thus $\delta^*_1>0$, $ y^*_{11} =  y^*_{12}=0$. The KKT variables are the same as the proof of Theorem \ref{thm:equiv_AV} in the situation where $s\leq(1-\beta)\omega$ and $x^{alt*}_2>0$. Without loss of generality, we  therefore conclude that 
\begin{align}
 s &= \frac{(1+\beta)(\alpha_{21}-\beta\alpha_{12})}{\alpha_{21}+\alpha_{12}}\cdot \omega\\
 &= \frac{(1-\beta)(\alpha_{21}+\beta\alpha_{12})}{\alpha_{21}+\alpha_{12}}\cdot \omega  \\
 &= \frac{1}{2}(1-\beta)\beta\omega
\end{align}
and $\alpha_{11} = \alpha_{22}$.

Now consider the possible optimal solutions
\begin{align}
\nonumber  x^*_1&=\beta [\alpha_{11} (d^*_1-z^*_1)+ \alpha_{21}(d^*_2-z^*_2)+y^*_{21}] + \delta^*_1\\
\nonumber  x^*_2&=\beta [\alpha_{12} (d^*_1-z^*_1)+ \alpha_{22}(d^*_2-z^*_2)] \\
\nonumber   x^*_1&= d^*_1 - z^*_1\\
\nonumber  y^*_{21} &=z^*_2 + x^*_2 - d^*_2\\
\nonumber  z^*_1&= z^*_2.
\end{align}
  
Suppose $d^*_1\geq d^*_2$. Then $d^*_1-z^*_1 \geq d^*_2-z^*_2$ and $x^*_2=\beta [\alpha_{12} (d^*_1-z^*_1)+ \alpha_{22}(d^*_2-z^*_2)] = \beta [\alpha_{21} (d^*_1-z^*_1)+ \alpha_{22}(d^*_2-z^*_2)] \geq \beta(d^*_2-z^*_2)$. Now let $z_1 = z_2 = d^*_2$ (increase both by $d^*_2-z^*_2$). Then decrease $x_2$ by $\beta(d^*_2-z^*_2)$ and $x_1$ by $(d^*_2-z^*_2)$, thus we decrease $\delta_1$ by $(1-\beta)(1+\beta)(d^*_2-z^*_2) < (1-\beta)(x^*_1+x^*_2)$. Hence we increase the cost by $2(d^*_2-z^*_2)\cdot s = 2(d^*_2-z^*_2)\cdot \frac{1}{2}(1-\beta)\beta\omega = (1-\beta) \beta(d^*_2-z^*_2) \cdot\omega$ and subsequently decrease the cost by $(1-\beta)(1+\beta)(d^*_2-z^*_2)\cdot\omega> (1-\beta) \beta(d^*_2-z^*_2) \cdot\omega$ (by \ref{eq:opt_a1}, $x_2>0$ indicates that $d^*_2-z^*_2>0$). Hence the total profit increases, which contradicts the fact that this is a profit-maximizing optimum.

Suppose $d^*_1< d^*_2$. With the same process as before, we increase $z_1$ and $z_2$ by $(d^*_1-z^*_1)$, decrease $x_2$ by $\beta(d^*_1-z^*_1)$ and $x_1$ by $d^*_1-z^*_1$, that is, we decrease $\delta_1$ by $(1-\beta)(1+\beta)(d^*_1-z^*_1) < (1-\beta)(x^*_1+x^*_2)$. Hence we increase the cost by $2(d^*_1-z^*_1)\cdot s = (1-\beta) \beta(d^*_1-z^*_1) \cdot\omega$ and subsequently decrease the cost by $(1-\beta)(1+\beta)(d^*_1-z^*_1)\cdot\omega> (1-\beta) \beta(d^*_1-z^*_1) \cdot\omega$, with the net effect of increasing the profit, which again is a contradiction.

Therefore $y_{ij} > 0$ is not an optimal solution in this situation.

\end{proof}

\end{proofcomment}

\change{The main result of this section below uses the above lemmas to establish that a profit-maximizing platform is able to realize the same optimal profits under either the HV priority or AV priority assignments.}

\begin{thm}
\label{thm: driver = AV}
Under Assumptions \ref{assum:all} and \ref{assum:obj_concave}, for any choice of $\omega, s,\beta$ and $\mathbf{A}$,  $\mathbf{u}^{*} = \left\{p_i^{*}, \delta_i^{*}, z_i^{*}, x_i^{*}, y_{ij}^{*}, r_{ij}^{*}\right\}_{i,j=1}^n$ is an optimal solution of the optimization problem for HV priority assignment \eqref{eq:opt_d1} if and only if it is an optimal solution of the optimization problem for AV priority assignment \eqref{eq:opt_a1}, and therefore the optimal profits of the two optimization problems are the same. 
\end{thm}

\begin{proofsketchcomment}
\begin{proofsketch}
With \change{Lemmas} \ref{prop: driver mix} and \ref{prop: AV mix}, we  obtain that, when the optimal solution for HV priority assignment \eqref{eq:opt_d1} and that for AV priority assignment \eqref{eq:opt_a1} are under the same deployment, i.e., both are HV-only deployment, mixed autonomy deployment or AV-only deployment, they must have the same optimal solutions. Theorem \ref{thm: driver = AV} follows by considering the optimal profits under these same deployment conditions.
\end{proofsketch}
\end{proofsketchcomment}

\begin{proofcomment}
\begin{proof}\

First notice that in each priority assignment, an optimal solution falls into one of three cases: HV-only (i.e., $z_i=0$ for all $i$), mixed autonomy (i.e., there exists some $i,j$ such that $x_i>0$ and $z_j>0$), and AV-only (i.e., $x_i=0$ for all $i$). In the case of HV-only or AV-only, it is straightforward to observe that when a solution is feasible for either HV priority assignment or AV priority assignment, it will also be feasible for the other AV assignment (consider the original optimization problems here). This is also true for the mixed case, since from Lemmas \ref{prop: driver mix} and \ref{prop: AV mix}, we know that $r_{ij} = y_{ij} = 0$ in both priority assignments. Therefore, the solutions for the two optimization problems are convertible: given $\beta, \omega, s$ and $A$, if a  solution is optimal for one priority assignment, it is also optimal for the other priority assignment. 

Since the objective functions of the two optimization problems \eqref{eq:opt_d1} and \eqref{eq:opt_a1} are the same, then the result above implies that they have the same optimal profits.
\end{proof}
\end{proofcomment}

We can then derive a threshold on the cost of AVs above which the platform does not find it optimal to deploy any AVs.

\begin{prop}
\label{prop:lowerbound-general}

Under Assumptions \ref{assum:all} and \ref{assum:obj_concave}, if $k> 1$, then, under any priority assignment, it is optimal for the platform to use an HV-only deployment, \emph{i.e.}, there is no benefit to introducing AVs into the ride-sharing network.
\end{prop}

\begin{proofcomment}
\begin{proof}\

Firstly we will develop another necessary condition.

Since we have proved that the two priority assignments achieve the same optimal solutions, then the following are equivalent:
\begin{itemize}
\item the inequality/equality in \eqref{eq:dk-1}/\eqref{eq:dk-2}/\eqref{eq:dk-3}/\eqref{eq:dk-4} holds 
\item the inequality/equality in \eqref{eq:ak-1}/\eqref{eq:ak-2}/\eqref{eq:ak-3}/\eqref{eq:ak-4} holds 
\item the inequality/equality in \eqref{eq:ak-5}/\eqref{eq:ak-6}/\eqref{eq:ak-7}/\eqref{eq:ak-8} holds. 
\end{itemize}

Moreover, consider the corresponding KKT condition for prices $p_i$, and denote the variables in \eqref{eq:dk-1}--\eqref{eq:dk-4} using superscript $d$. The KKT conditions require $\frac{\partial (p_i d_i)}{\partial p_i}(p^*_i)+ \frac{\partial d_i }{\partial p_i}(p^*_i)(\sum_j \alpha_{ij}\mu^d_j - \gamma^d_i) = \frac{\partial (p_i d_i)}{\partial p_i}(p^*_i)+ \frac{\partial d_i }{\partial p_i}(p^*_i)(\sum_j \alpha_{ij} \beta\lambda_j - \gamma_i) = \frac{\partial (p_i d_i)}{\partial p_i}(p^*_i)+ \frac{\partial d_i }{\partial p_i}(p^*_i)(\sum_j \alpha_{ij} \beta\lambda^1_j - \gamma^1_i) = 0$. The last equality holds because $p^*_i>0$ for all $i$ obviously. Hence $\sum_j \alpha_{ij}\mu^d_j - \gamma^d_i = \sum_j \alpha_{ij} \beta\lambda_j - \gamma_i = \sum_j \alpha_{ij} \beta\lambda^1_j - \gamma^1_i$.

Therefore, satisfying the relation of \eqref{eq:dk-1}--\eqref{eq:dk-4} with \eqref{eq:ak-5}--\eqref{eq:ak-8} requires $-\omega +\lambda^d_i = -\omega +\lambda^1_i$, $\sum_j \alpha_{ij}(\beta\lambda^d_j - \mu^d_j) -\lambda^d_i+\gamma^d_i = -\lambda^1_i+\gamma^1_i$, $-s+\gamma^d_i - \mu^d_i = -s -\beta \lambda^1_i + \sum_j \alpha_{ij} \mu^1_j + \gamma^1_i-\mu^1_i$ and $ \mu^d_j - \gamma^d_i = \beta \lambda^1_j - \gamma^1_i $ for all $i,j$.

These requirements yield that $\lambda^d_i = \lambda_i^1$ and $\gamma^1_i = \gamma_i^d+c$ where $c = \beta \lambda^d_j-\mu^d_j$ for any $j$. In addition, 
\begin{align*}
	-s + \gamma^d_i-\mu^d_i &= -s -\beta \lambda^1_i + \sum_j \alpha_{ij} \mu^1_j + \gamma^1_i-\mu^1_i\\
	\gamma^d_i-\mu^d_i &= -\beta \lambda^1_i + \sum_j \alpha_{ij} \mu^1_j + \gamma^1_i-\mu^1_i\\
	\gamma^d_i-\mu^d_i &= -\beta \lambda^1_i + \sum_j \alpha_{ij} \mu^1_j+(\gamma_i^d+\beta \lambda^d_i-\mu^d_i) - \mu^1_i\\
	 0 &=  \sum_j \alpha_{ij} \mu^1_j - \mu^1_i
\end{align*} and applying this to \eqref{eq:ak-7} gives a new necessary condition that must be satisfied for any optimal solution for the optimization problem \eqref{eq:opt_a2}:
\begin{equation}
\label{eq:17}
-s-\beta\lambda^1_i+\gamma^1_i\leq 0
\end{equation}
where the equality holds when $z_i>0$.

With the condition described in \eqref{eq:17} held, we can construct the threshold for the cost of AV above which the mixed-autonomy won't be beneficial for the platform.

Assume the optimal profit of the mixed autonomy deployment is strictly greater than that of the HV-only deployment. Then there exists a location $i$ such that $z_i>0$. Hence by \eqref{eq:16}, $-s-\beta\lambda^1_i+\gamma^1_i = 0$. Moreover, from \eqref{eq:ak-4}, \eqref{eq:ak-5} and \eqref{eq:ak-8}, $\beta\lambda^1_j \leq \gamma^1_i \leq \lambda^1_i \leq \omega$ for any $j$. Therefore, $s = \gamma^1_i -\beta\lambda^1_i \leq \lambda^1_i - \beta\lambda^1_i\leq (1-\beta)\omega$. Hence $k = \frac{s}{\omega} \leq \frac{(1-\beta)\omega}{\omega} = 1-\beta$.

\end{proof}
\end{proofcomment}

\change{Before presenting and analyzing a third priority assignment, we discuss restrictions of the present model which posits several simplifying assumptions such as equidistant locations. First, such assumptions might be reasonable in certain settings. For example, about 75\% of taxi rides in New York City are less than three miles\footnote{As determined from almost 7 million yellow taxi trips in June 2019 available at https://www1.nyc.gov/site/tlc/about/tlc-trip-record-data.page}, suggesting that distance may not be a major distinguishing attribute of most rides in that market. Moreover, \cite{Bimpikis2018} includes discussion on how to potentially relax such assumptions. Second, even with these simplifications, the theoretical analysis and results presented here are challenging, suggesting that a full treatment in a more general setting is difficult and motivating first a thorough study in a simplified setting. Lastly, simplifying assumptions allow for fundamental insights such as in Theorem \ref{thm: driver = AV} and below in Section \ref{sec:numerical-study} that are not obscured or confounded by additional degrees of freedom.}

\section{Weighted Priority Assignment}
\label{sec:weighted-model}
Besides assigning the rides to one type of vehicle---HVs or AVs---first, and then using the other type to satisfy any remaining demand, it is also reasonable to consider that any vehicle in the platform can be chosen randomly with equal probability. Therefore, in this section, we  introduce the \emph{weighted priority assignment} in which the platform assigns the rides at each location to HVs and AVs at that location with the same probability, \emph{i.e.}, in proportion to the relative fraction of HVs and AVs to the total number of vehicles.

\subsection{Equilibrium Definition for Weighted Priority Assignment}

As described above, in weighted priority assignment, HVs and AVs are assigned to riders with equal possibility: $Prob\left\{\text{rider assigned to HV} \right\} = Prob\left\{\text{rider assigned to AV} \right\} = \min\{\frac{\theta_i(1-F(p_i))}{x_i +z_i},1\}$ for all $i$. The resulting equilibrium constraints for the model are:
\begin{align}
\label{eq:12}\nonumber x_i &=\beta \Big[\sum_{j} \alpha_{ji}  \min \left\{ 1 ,    \frac{\theta_j(1-F(p_j))}{x_j +z_j}\right\}\cdot x_j \\
 & \qquad + \sum_{j} y_{ji} \Big] + \delta_i\\
 \label{eq:13}
\sum_j y_{ij}&= \max\left\{ 1 -   \frac{\theta_i(1-F(p_i))}{x_i +z_i} , 0  \right\}\cdot x_i\\
\label{eq:14}
z_i&=\sum_{j} \alpha_{ji} \min \left\{ 1 , \frac{\theta_j(1-F(p_j))}{x_j +z_j}\right\}\cdot z_j + \sum_{j} r_{ji}\\
\label{eq:15}
\sum_j r_{ij} &= \max \left\{0,1- \frac{\theta_i(1-F(p_i))}{x_i +z_i}  \right\}\cdot z_i.
\end{align}

The expected lifetime earnings $V_i$ for a driver at location $i$ takes the form
\begin{align}
\nonumber V_i &= \min \left\{\frac{\theta_i(1-F(p_i))}{x_i +z_i}, 1 \right\}(c_i + \sum_{k=1}^n\alpha_{ik}\beta V_{k}) \\
\label{eq:16}  &+ \left(1-\min \left\{\frac{\theta_i(1-F(p_i))}{x_i +z_i}, 1 \right\}\right)\beta\max_j V_j.
\end{align}

\change{As before, the platform chooses compensation such that $V_i = \omega$.}

\begin{definition}
For some prices and compensations $\{p_i,c_i\}_{i=1}^n$, the collection $\left\{\delta_i,x_i,y_{ij},z_i,r_{ij} \right\}_{i,j=1}^n$ is \emph{an equilibrium under $\{p_i,c_i\}_{i=1}^n$ for weighted priority assignment} if \eqref{eq:12}--\eqref{eq:15} is satisfied and $V_i$ as defined in \eqref{eq:16} satisfies $V_i=\omega$ for all $i= 1,\ldots, n$ \change{such that $\delta_i+\sum_{j=1}^n y_{ji}>0$.}
\end{definition}

To further study weighted priority assignment, we now introduce the following assumption which ensures that the platform can make some profit by offering rides at an appropriate price. 
\change{
\begin{assum}
\label{assum:concave-d}
The parameters $\beta, \omega$ and $s$ are such that $(1-\beta)\omega < \bar{p}$ or $s<\bar{p}$.
\end{assum}
}

\subsection{Profit-Maximization Optimization Problem for Weighted Priority Assignment}
We now establish the following profit-maximization problem for weighted priority assignment:
\begin{align}
\nonumber  \max_{\{p_i,c_i\}_{i=1}^n}&\sum_{i=1}^n \left[ \min \left\{x_i+z_i, \theta_i (1-F(p_i)) \right\} \cdot p_i \right. \\
\nonumber & \left. - \min \left\{x_i, \theta_i(1-F(p_i)) \frac{x_i}{x_i +z_i} \right\} \cdot c_i - z_i\cdot s\right]  \\
\nonumber \text{s.t.} &\left\{\delta_i,x_i,y_{ij},z_i,r_{ij} \right\}_{i,j=1}^n \text{ is an equilibrium}\\
\label{eq:opt_w0} & \text{ under } \{p_i,c_i\}_{i=1}^n\text{ for weighted priority assignment}.
\end{align}

As in Section \ref{sec:optimization-problem}, we establish an equivalent optimization problem 
\begin{align}
\nonumber \max_{\{p_i,\delta_i,x_i,y_{ij},z_i,r_{ij}\}}&\sum_{i=1}^n p_i \theta_i (1-F(p_i)) - \omega\sum_{i=1}^n \delta_i - s\sum_{i=1}^n z_i\\
\nonumber \text{s.t.}\quad d_i=&\theta_i (1-F(p_i))  \\
\nonumber  x_i=&\beta \left[\sum_{j} \alpha_{ji} d_j \frac{x_j}{x_j +z_j}+ \sum_{j} y_{ji} \right] + \delta_i\\
\nonumber \sum_{j=1}^n y_{ij}=& x_i -d_i\frac{x_i}{x_i +z_i}   \\
\nonumber  z_i=&\sum_{j=1}^n \alpha_{ji}d_j\frac{z_j}{x_j +z_j} + \sum_{j=1}^n r_{ji}\\
\nonumber \sum_{j=1}^n r_{ij}=& z_i - d_i \frac{z_i}{x_i +z_i} \\
  \label{eq:opt_w1} p_i,&\delta_i,z_i,x_i,y_{ij},r_{ij}\geq 0 \qquad \forall i,j,
\end{align}
 followed by a lemma showing the equivalence.

\begin{lemma} 
\label{lem:opt_w021}
\change{Assume weighted priority assignment and consider the optimization problems \eqref{eq:opt_w0} and \eqref{eq:opt_w1}. Under Assumptions \ref{assum:all}, \ref{assum:obj_concave} and \ref{assum:concave-d}, an optimal solution to \eqref{eq:opt_w1} provides an optimal solution to \eqref{eq:opt_w0}. In particular, any optimal solution $\left\{p_i^*, \delta^*_i, x_i^*, y_{ij}^*, z_i^*, r_{ij}^* \right\}$ for \eqref{eq:opt_w1} is such that $d_i^*>0$ for all $i$, \emph{i.e.}, some riders are served at all locations, and there exist compensations $\left\{c^*_i \right\}_{i=1}^n$ such that $\left\{\delta^*_i, x^*_i, y^*_{ij}, z^*_i, r^*_{ij} \right\}_{i,j=1}^n$ constitutes an equilibrium under $\left\{p^*_i, c^*_i \right\}_{i=1}^n$ for weighted priority assignment. Moreover, $\left\{p^*_i, c^*_i \right\}_{i=1}^n$ is optimal for \eqref{eq:opt_w0}.}

\end{lemma}

\begin{proofcomment}
\begin{proof}\

The proof for the first two points are similar to that of the HV and AV priority assignments. Obviously, $d^*_i \leq x^*_i +z^*_i$ for \eqref{eq:opt_w0}. Hence we can turn the equilibrium constraints into the constraints in \eqref{eq:opt_w1}. By setting the compensation $c_i = \omega(1-\beta)\cdot \frac{x_i+z_i}{d_i}$ for all $i$, we  obtain the equivalent optimization \eqref{eq:opt_w1}.

Consider the optimization problem \eqref{eq:opt_w1} of weighted priority assignment and compare it with that of HV priority assignment \eqref{eq:opt_d1}. By observation, if for any optimal solution of HV priority assignment, we can obtain that $\min\left\{x^*_i,d^*_i\right\} = d^*_i \cdot \frac{x^*_i}{x^*_i + z^*_i}$ and  $\max\left\{d^*_i - x^*_i,0\right\} = d^*_i \cdot \frac{z^*_i}{x^*_i + z^*_i}$ (notice that $\max\left\{ x^*_i - d^*_i,0\right\} = x_i -  \min\left\{x^*_i,d^*_i\right\}$), then it follows that any optimal solution for HV priority assignment will be feasible for weighted priority assignment. 

By Assumption \ref{assum:concave-d}, we have that $(1-\beta)\omega<\bar{p}$ or $s<\bar{p}$. Hence Lemma \ref{lem:opt_d021} establishes that $x^*_i + z^*_i\geq d^*_i>0$ for all $i$. We then consider the optimal solution in the three cases.

If it falls in the HV-only case, i.e., $x^*_i > 0, z^*_i=0$ for all $i$, then this implies $d^*_i \leq x^*_i$ for all $i$. Therefore, we have
\begin{align*}
  \begin{cases}
                  d^*_i \cdot \frac{x^*_i}{x^*_i + z^*_i} =  d^*_i & = \min\left\{x^*_i,d^*_i\right\}\\
                  d^*_i \cdot \frac{z^*_i}{x^*_i + z^*_i} =  0 & = \max\left\{d^*_i - x^*_i,0\right\} .
\end{cases}
\end{align*}

Similarly, if the optimal solution is in the AV-only case, i.e., $x^*_i = 0, z^*_i>0$, then $d^*_i \geq x^*_i$ for all $i$. Hence 
\begin{align*}
  \begin{cases}
                  d^*_i \cdot \frac{x^*_i}{x^*_i + z^*_i} =  0 & = \min\left\{x^*_i,d^*_i\right\}\\
                  d^*_i \cdot \frac{z^*_i}{x^*_i + z^*_i} =  d^*_i & = \max\left\{d^*_i - x^*_i,0\right\} .
\end{cases}
\end{align*}

Lastly, when the optimal solution is in the mixed autonomy case, i.e., $x^*_i > 0, z^*_i>0$ for some $i$, then $d^*_i \geq x^*_i$ for all $i$. Also, Proposition \ref{prop: driver mix} implies that $y^*_{ij} = r^*_{ij}=0$ here for all $i,j$, and then $d^*_i = x^*_i +z^*_i $ for all $i$. Therefore, we  observe that 
\begin{align*}
  \begin{cases}
                  d^*_i \cdot \frac{x^*_i}{x^*_i + z^*_i} =  x^*_i & = \min\left\{x^*_i,d^*_i\right\}\\
                  d^*_i \cdot \frac{z^*_i}{x^*_i + z^*_i} =  d^*_i - x^*_i & = \max\left\{d^*_i - x^*_i,0\right\} .
\end{cases}
\end{align*}

Thus, the optimal solutions for the HV and AV priority assignments are always feasible for weighted priority assignment. Hence, under Assumption \ref{assum:concave-d}, any optimal solution $\left\{p_i^*, \delta^*_i, x_i^*, y_{ij}^*, z_i^*, r_{ij}^* \right\}$ for \eqref{eq:opt_w1} is such that $d_i^*>0$ for all $i$. 

\end{proof}
\end{proofcomment}

The following theorem establishes that weighted priority assignment obtains the same optimal profits as the HV and AV priority assignments, which were already shown to obtain the same optimal profits in Theorem \ref{thm: driver = AV}.

\begin{thm}
\label{thm: Weighted = Prior}
Under Assumptions \ref{assum:all}, \change{\ref{assum:obj_concave}} and \ref{assum:concave-d}, for any choice of $\omega, s,\beta$ and $\mathbf{A}$, a feasible solution $\mathbf{u}$ for \eqref{eq:opt_d1} or  \eqref{eq:opt_a1} is optimal for \eqref{eq:opt_d1} or \eqref{eq:opt_a1}  if and only if $\mathbf{u}$ is an optimal solution for $ \eqref{eq:opt_w1}$.
\end{thm}

\begin{proofsketchcomment}
\begin{proofsketch}
By recombining the constraints in \eqref{eq:opt_w1}, we can obtain another optimization problem given by
\begingroup
\allowdisplaybreaks
\begin{align}
\nonumber \max_{\{p_i,\delta_i,x_i,y_{ij},z_i,r_{ij}\}}&\sum_{i=1}^n p_i \theta_i (1-F(p_i)) - \omega\sum_{i=1}^n \delta_i - s\sum_{i=1}^n z_i\\
\nonumber \text{s.t.}\quad d_i=&\theta_i (1-F(p_i))  \\
\nonumber  x_i + \beta z_i=&\beta \left[\sum_{j} \alpha_{ji} d_j + \sum_{j} y_{ji}+ \sum_{j} r_{ji} \right] + \delta_i\\
\nonumber \sum_{j=1}^n y_{ij}+ \sum_{j=1}^n r_{ij}=& x_i+z_i -d_i \\
\nonumber z_i -  \sum_{j=1}^n \alpha_{ji} z_j= & \sum_{j=1}^n r_{ji} - \sum_{j=1}^n \alpha_{ji} \sum_{k=1}^n r_{jk} \\
  \label{eq:opt_w2} p_i,\delta_i,&z_i,x_i,y_{ij},r_{ij}\geq 0 \qquad \forall i,j.
\end{align}
\endgroup
By construction, any optimal solution for \eqref{eq:opt_w1} will be feasible for \eqref{eq:opt_w2} and thus the optimal profit of \eqref{eq:opt_w2} will be no less than that of \eqref{eq:opt_w1}.

By \change{Assumptions \ref{assum:obj_concave} and} \ref{assum:concave-d}, \eqref{eq:opt_w2} is a convex optimization problem with affine constraints, and thus the KKT conditions are not only necessary, but also sufficient for optimality. By studying the KKT conditions, we conclude any optimal solution for AV priority assignment is also optimal (and feasible) for \eqref{eq:opt_w2}. since the optimal profits for \eqref{eq:opt_w2} are higher than or equal to that of \eqref{eq:opt_w1}, and since any optimal solution for AV priority assignment is feasible for \eqref{eq:opt_w1}, then we can conclude that any optimal solution for AV priority assignment is also optimal (and feasible) for \eqref{eq:opt_w1}.    
\end{proofsketch}
\end{proofsketchcomment}

\begin{proofcomment}
\begin{proof}\

By recombining the constraints in \eqref{eq:opt_w1}, we can obtain another optimization problem given by

\begin{align}
\nonumber \max_{\{p_i,\delta_i,x_i,y_{ij},z_i,r_{ij}\}}&\sum_{i=1}^n p_i \theta_i (1-F(p_i)) - \omega\sum_{i=1}^n \delta_i - s\sum_{i=1}^n z_i\\
\nonumber \text{s.t.}\quad d_i=&\theta_i (1-F(p_i))  \\
\nonumber  x_i + \beta z_i=&\beta \left[\sum_{j} \alpha_{ji} d_j + \sum_{j} y_{ji}+ \sum_{j} r_{ji} \right] + \delta_i\\
\nonumber \sum_{j=1}^n y_{ij}+ \sum_{j=1}^n r_{ij}=& x_i+z_i -d_i \\
\nonumber z_i -  \sum_{j=1}^n \alpha_{ji} z_j= & \sum_{j=1}^n r_{ji} - \sum_{j=1}^n \alpha_{ji} \sum_{k=1}^n r_{jk} \\
  \label{eq:opt_w2} \delta_i,&z_i,x_i,y_{ij},r_{ij}\geq 0 \qquad \forall i,j.
\end{align}
By construction, any optimal solution for \eqref{eq:opt_w1} will be feasible for \eqref{eq:opt_w2} and thus the optimal profit of \eqref{eq:opt_w2} will be no less than that of \eqref{eq:opt_w1}.

As we have already proved in Lemma \ref{lem:opt_w021}, the optimal solution of the optimization problem in priority assignment is always a feasible solution for \eqref{eq:opt_w1}.

Consider the optimization problem \eqref{eq:opt_w2}, and rewrite it by considering $d_i$ as the variable instead of $p_i$. Notice that since $d_i = \theta_i (1-F(p_i))$ is monotonically decreasing, we are able to write $p_i$ as a function $d_i$ because the inverse mapping exists. Moreover, we can relax the constraint $p_i \geq 0$ for all $i$ since $d_i$ is always positive and thus a negative price cannot be optimal. 

Below lists the KKT conditions related to \eqref{eq:opt_w2} while regarding $d_i$ as a variable instead of $p_i$: 
\begin{align}
  \label{eq:wk-1} \text{(constraints on $d_i$)}: & \frac{\partial{p_i d_i}}{\partial{d_i}}+ \beta \sum_j \alpha_{ij}  \lambda_j - \gamma_i = 0\\
  \label{eq:wk-2} \text{(constraints on $\delta_i$)}: & -\omega +\lambda_i \leq 0\\ 
  \label{eq:wk-3} \text{(constraints on $x_i$)}: & -\lambda_i+\gamma_i\leq 0 \\
  \label{eq:wk-4} \text{(constraints on $z_i$)}: & -s - \sum_j \alpha_{ij} (\beta \lambda_j-\mu_j) + \gamma_i-\mu_i\leq 0\\
  \label{eq:wk-5} \text{(constraints on $y_{ij}$)}: & \beta \lambda_j-\gamma_i \leq 0\\
  \label{eq:wk-6}  \text{(constraints on $r_{ij}$)}: & \beta \lambda_j-\gamma_i - \mu_j +  \sum_j \alpha_{ij}\mu_j \leq 0.
\end{align}

By Assumption \ref{assum:obj_concave} and \ref{assum:concave-d}, \eqref{eq:opt_w2} is a convex optimization problem with affine constraints, and thus the KKT conditions are not only necessary, but also sufficient for optimality. Hence in order to show a solution to be optimal for \eqref{eq:opt_w2}, it is enough to show that it satisfies all the KKT conditions \eqref{eq:wk-1}--\eqref{eq:wk-6}:

Given the optimal solution and the KKT variables $\lambda^1_i$ and $\gamma^1_i$ resolved from the optimal solution of AV priority assignment with the conditions \eqref{eq:ak-5}--\eqref{eq:ak-8}, let $\mu_i = \mu_j$ for all $i,j$. Then the conditions \eqref{eq:wk-1}--\eqref{eq:wk-6} and the constraints for weighted priority assignment can all be satisfied. Therefore, any optimal solution for AV priority assignment is also optimal (and feasible) for \eqref{eq:opt_w2}. 

At the same time, since the optimal profits for \eqref{eq:opt_w2} are higher than or equal to that of \eqref{eq:opt_w1}, and since any optimal solution for AV priority assignment is feasible for \eqref{eq:opt_w1}, then we can conclude that any optimal solution for AV priority assignment is also optimal (and feasible) for \eqref{eq:opt_w1}. \end{proof}
\end{proofcomment}

\change{Theorems \ref{thm: driver = AV} and \ref{thm: Weighted = Prior} show that, even though the three  priority assignments prescribe different models for incorporating AVs into a ride-sharing platform, the resulting profits at an optimal equilibrium are the same in all three cases under Assumptions \ref{assum:all}, \change{\ref{assum:obj_concave}} and \ref{assum:concave-d}. This is because no location will have both AVs and HVs present at an optimal equilibrium. Intuitively, on the one hand, the platform is able set compensation for drivers and to deploy AVs as desired, so that there is considerable freedom in dictating system operation. On the other hand, locations are coupled through the rider demand pattern and cannot be managed independently by the platform, highlighting the surprising nature of this result.}

\section{Closed-Form Characterization for Star-to-Complete Networks}
\label{sec:numerical-study}
In this section, we consider the family of \emph{star-to-complete networks} introduced in \cite{Bimpikis2018}. \change{For this large class of networks, we derive closed form expressions for the thresholds of relative cost between HVs and AVs for which the platform finds it optimal to use an HV-only deployment, AV-only deployment, or a mixed autonomy deployment.}

\begin{definition}
The class of demand patterns $(\mathbf{A}^\xi,\mathbf{1})$ with $n\geq 3$, $\xi \in \left[0,1 \right]$, and
\begin{align}
\mathbf{A}^\xi &= \begin{bmatrix}
    0 & \frac{1}{n-1} & \frac{1}{n-1} & \dots & \frac{1}{n-1} \\
    c_1 & 0 & c_2 & \dots & c_2 \\   
    c_1 & c_2 & 0 & \dots & c_2 \\
    \vdots &  \vdots &  \vdots &  \ddots &  \vdots\\
    c_1 & c_2 & \dots & c_2 & 0 \\
\end{bmatrix},\\
c_1 &= \frac{\xi}{n-1}+(1-\xi),\qquad
c_2 = \frac{\xi}{n-1}
\end{align}
is the family of \emph{star-to-complete} networks. It is a \emph{star} network when $\xi=0$ for which we write $\mathbf{A}^S=\mathbf{A}^0$ and a \emph{complete} network when $\xi=1$ for which we write $\mathbf{A}^C=\mathbf{A}^1$. Therefore the general adjacency matrix of a \emph{star-to-complete} network can be written as $\mathbf{A}^\xi$ = $\xi \mathbf{A}^{C} + (1-\xi) \mathbf{A}^{S}$.
\end{definition}

In addition, we make the following assumption throughout this section.
\begin{assum}
\label{assum:2}
  All locations have the same mass of potential riders, which we normalize to one, i.e., $\mathbf{\theta}=\mathbf{1}$. Also, the riders' willingness to pay is uniformly distributed in $\left[0,1 \right]$ so that $F(p)=p$ for $p\in[0,1]$.
\end{assum}

Consider fixed outside option earnings $\omega$, and recall the parameter $k$ determining the  cost of operating AVs for the same lifetime of an HV relative to $\omega$. In this section, we confirm the intuition that, for large $k$, i.e. high relative cost of AVs, the profit maximizing strategy for the platform is an HV-only deployment, and for small $k$, i.e. low relative cost of AVs, the profit maximizing strategy for the platform is an AV-only deployment. We also show that in some cases, but not all, for some values of $k$, the platform finds it optimal to use both HVs and AVs at equilibrium, \emph{i.e.}, a true mixed autonomy deployment. 

Recall that Proposition \ref{prop:lowerbound-general} provides a sufficient condition for when a platform will not find it optimal to use AVs. In the next Theorem, we sharpen this result for the class of star-to-complete networks and fully characterize the regions in which the profit-maximizing platform will deploy an HV-only deployment, an AV-only deployment, and a truly mixed autonomous network.

\change{
\begin{thm}
\label{thm:closed_form}
Consider a star-to-complete network under Assumption \ref{assum:2}. Define
\begin{align*}
  k_1 &= \frac{1+\beta c_1}{c_1 + 1} \, ,\\
  k_2 &= \begin{cases}
                  1 &\text{if }\xi \in [\frac{\beta(n-1)-1}{\beta(n-2)},1]\\
                  \frac{c_1(1+\beta) + (n-1)\beta^2 c_1^3 + 1}{(c_1 + 1)((n-1)\beta^2 c_1^2 + 1)} & \text{if } \xi \in [\beta_{lim} ,\frac{\beta(n-1)-1}{\beta(n-2)})\\
                  \frac{1+\beta c_1}{c_1 + 1} & \text{if }\xi \in [0,\beta_{lim}),
\end{cases}\\
  k_3 &= \frac{(n-1)c_1-1}{(1-\beta)(n-1)(1+c_1)c_1}\, ,\\
  k_4 &= \frac{(1+\beta)c_1 + (n-1)\beta c_1^3 + 1}{(c_1 + 1)(\beta(n-1) c_1^2 + 1)} \, ,
\end{align*}
where 
\begin{align*}
 \beta_{lim} = \max\bigg\{ &\frac{n-1}{2(1-\beta)\beta(n-2)}\bigg[\beta(1-2\beta)\\
 &+\sqrt{\frac{\beta^2(n-1)+4\beta-4}{n-1}}\bigg],0\bigg\}.
\end{align*}

Suppose $k_3 \geq k_1$, equivalently, $\beta c_1(n-1)(1-c_1+\beta c_1) \geq 1$. When $k\in[0,k_1]$, it is always optimal for the platform to deploy an AV-only deployment, \emph{i.e.}, optimal profits are obtained with $x_i=0$ for all $i$. If $k_1<k_2$, then: when $k\in(k_1,k_2)$, it is optimal for the platform to deploy a mixed autonomous network, \emph{i.e.}, optimal profits are obtained with $x_i>0$ and $z_j>0$ for some $i,j$; when $k\geq k_2$, it is optimal for the platform to deploy an HV-only deployment, \emph{i.e.}, optimal profits are obtained with $z_i=0$ for all $i$. If $k_1\geq k_2$, then: when $k> k_1$, it is optimal for the platform to deploy an HV-only deployment.

Now suppose  $k_3 < k_1$, equivalently, $\beta c_1(n-1)(1-c_1+\beta c_1) \geq 1$. When $k \in [0,k_4]$, it is optimal for the platform to deploy an AV-only deployment; when $k \in (k_4,k_2)$, it is optimal to deploy a mixed autonomy deployment; when $k \geq k_2$, it is optimal to deploy an HV-only deployment.

\end{thm}
}

\begin{proofcomment}
\begin{proof}
To prove the theorem, we only need to find the optimal solutions of the optimization problems \eqref{eq:opt_d0}, \eqref{eq:opt_a0} or \eqref{eq:opt_w0}, divide $k$ values into different regions according to the optimal solutions and find the intersection of $k$ values between different regions.

For convenience, we can first divide the optimal solutions into four possible regions:
\begin{enumerate}
\item HV only: $x_i > 0$, $z_j = 0$ for some $i$ and all $j$,
\item Mixed-autonomy: $x_i>0$, $z_i>0$ for some $i$.
\item AV only without transition: $x=0$, $z>0$, $r = 0$.
\item AV only with transition: $x=0$, $z>0$, $r_{ij} > 0$ for some $i,j$.
\end{enumerate}

\underline{Region 1:}

Notice that in this region, since $z_j = 0$ for all $j$, then $d_j \leq x_j$ for all $j$. As we've shown in the proof of Theorem \ref{thm:equiv_driver}, the optimal solutions for the first region with only HVs can thus be derived from \cite{Bimpikis2018} by setting $z_i = r_{ij} = 0$ for all $i,j$.

Let 
\begin{align*}
 \beta_{lim} = \max\bigg\{ &\frac{n-1}{2(1-\beta)\beta(n-2)}\bigg[\beta(1-2\beta)\\
 &+\sqrt{\frac{\beta^2(n-1)+4\beta-4}{n-1}}\bigg],0\bigg\}.
\end{align*}
$\beta_{lim}>0$ when $ (n-1)^{-\frac{1}{3}} < \beta < 1$.

\begin{enumerate}
    \item $\xi \in [\frac{\beta(n-1)-1}{\beta(n-2)},1]$: 
    
    $p_i = 1-\frac{\beta}{2}$, $x_i = 1-p_i=\frac{\beta}{2}$, $y_{ij}=0$ for all $i,j$; $\delta_1 = \frac{\beta-(n-1)\beta^2}{2}$, $\delta_i = \frac{(n-1)\beta-\beta^2}{4}$ for $i>1$, 
  
    \item $\xi \in [\max\left\{\beta_{lim},0\right\},\frac{\beta(n-1)-1}{\beta(n-2)})$:
    
    Let $Z = -(n-1)c_1$. Then
    \begin{equation*}
        p_i = \frac{1}{2}+\frac{\beta Z(1+\beta Z+\beta)+(n-1)[1-\beta c_2(n-2)]}{2(n-1)+2\beta^2 Z^2}
    \end{equation*}
    for $i>1$, and $p_1 = 1-\beta c_1(n-1)(1-p_2)$.
    
    $x_i = 1-p_i$, $y_{ij}=0$ for all $i,j$; 
    $\delta_1 = 0$, $\delta_i = (1-\beta c_2-\beta^2 c_1)(1-p_i)$ for $i>1$.

    \item $\xi \in \big[0,\max\left\{\beta_{lim},0\right\} \big)$:
    
    $p_1 = \frac{1}{2}$ and $p_i = \frac{1}{2}+ \frac{1-\beta^2 c_1-\beta(n-2)c_2}{2}$ for $i>1$. 
    
    $x_i = \sum_j y_{ij} + (1-p_i)$ for all i. $y_{1j} = \beta c_1(1-p_2)-\frac{1}{2(n-1)}$ for all $j>1$ and $y_{ij} = 0$ for all $i>1$ and all $j$.
    
    $\delta_1 = 0$, $\delta_i = (1-p_2)(1-\beta c_2-\beta^2 c_1)$ for all $i>1$.

\end{enumerate}

If $\beta \leq \frac{1}{n-1}$, then only the first case exists;
if $\frac{1}{n-1} < \beta \leq (n-1)^{-\frac{1}{3}}$, then only first two cases exist;
if $(n-1)^{-\frac{1}{3}} < \beta < 1$, then all three cases exist.

For the rest of regions, $z_i > 0$ for some $i$ and given Lemma \ref{prop: driver mix}, we can ensure that $y_{ij} =0$ for all $i, j$. By Theorem \ref{thm:equiv_driver}, it is without loss of generality for us to compute only the optimal solutions for \eqref{eq:opt_d2} knowing $y_{ij} =0$ for all $i, j$. Moreover, under Assumption \ref{assum:2}, the optimization problem \eqref{eq:opt_d2} become a quadratic problem with linear constraints. As a result, the KKT conditions are both necessary and sufficient for optimal solutions. Therefore, we first rewrite the simplified optimization problem of \eqref{eq:opt_d2} under Assumption \ref{assum:2} as below

\begin{align}
\nonumber \max_{\{p_i,\delta_i,x_i,z_i,r_{ij}\}}&\sum_{i=1}^n p_i (1-p_i) - \omega\sum_{i=1}^n \delta_i - s\sum_{i=1}^n z_i\\  
\nonumber  \text{s.t.}\quad x_i&=\beta \sum_{j=1}^n \alpha_{ji}x_j + \delta_i\\
\nonumber  z_i&=\sum_{j=1}^n \alpha_{ji} (1-p_j-x_j)+ \sum_{j=1}^n r_{ji}\\
\nonumber \sum_{j=1}^n r_{ij}&=z_i -(1-p_i-x_i )\\
\label{eq:opt_c} \delta_i&,x_i,z_i,r_{ij}\geq 0 \qquad \forall i,j.
\end{align}

We denote the dual variables for the three equality constraints as $\lambda_i, \mu_i$ and $\gamma_i$, for all $i = 1,\cdots, n$; $a_i, b_i, c_i$ and $d_{ij}$ for all $i,j = 1,\cdots, n$ are used to denote the four inequality constraints $-\delta_i, -x_i, -z_i, -r_{ij}\leq 0$. Therefore, the KKT conditions for \eqref{eq:opt_c} can be written as:

\begingroup
\allowdisplaybreaks
\begin{align}
    \label{eq:k1}   1-2p_i-\sum_{j=1}^n \alpha_{ij}\mu_j-\gamma_i &= 0\\
    \label{eq:k2}   -\omega+\lambda_i+b_i &= 0\\
    \label{eq:k3}   \sum_{j=1}^n(\beta \lambda_j - \mu_j) - \lambda_i + c_i &= 0\\
    \label{eq:k4}   -s - \mu_i - \gamma_i + a_i &= 0 \\
    \label{eq:k5}   \mu_j + \gamma_i +d_{ij} &= 0\\
    \label{eq:k6}   \lambda_i (\beta \sum_{j=1}^n \alpha_{ji}x_j + \delta_i - x_i) &= 0\\  
    \label{eq:k7}   \mu_i \Big(\sum_{j=1}^n \alpha_{ji} (1-p_j-x_j)+ \sum_{j=1}^n r_{ji}-z_i \Big) &= 0\\  
    \label{eq:k8}   \gamma_i \Big(\sum_{j=1}^n r_{ij}-z_i + (1-p_i-x_i ) \Big) &= 0\\  
    \label{eq:k9}   a_i z_i = b_i \delta_i = c_i x_i = d_{ij} r_{ij} &=0\\
    \label{eq:k10}  a_i, b_i, c_i, d_{ij} &\geq 0  \quad \forall i,j.
\end{align}
\endgroup

Solving equations \eqref{eq:k1}--\eqref{eq:k10} provides us an optimal solution for \eqref{eq:opt_c} and thus for \eqref{eq:opt_d1}. 

\underline{Region 2:}
\begin{align}
     \nonumber p_1 &= \frac{k(c_1+1)+(1-\beta)c_1-1}{2c_1},\\
     \nonumber p_i &= 1-\frac{\beta k(c_1+1)+\beta c_1}{2} \quad \forall i>1.
\end{align}

$x_i = \frac{(n-1)c_1(1-p_2)-(1-p_1)}{(n-1)(1-\beta)c_1}$ for all $i>1$, $x_1 =c_1(n-1)\beta x_2$.

$r_{ij} = 0$, $z_i = 1-p_i-x_i$ for all $i,j$.

$\delta_1=0$, $\delta_i = (1-\beta(1-c_1))x_2- \frac{\beta x_1}{n-1}$ for $i>1$.

\underline{Region 3:}
\begin{align}
     \nonumber p_i &= \frac{2(n-1)c_1^2+c_1(k(1-\beta)-1)+k(1-\beta)+1}{2[(n-1)c_1^2+1]}\\
     \nonumber &\qquad \forall i>1,\\
     \nonumber p_1 &= 1-(n-1)c_1(1-p_2) .
\end{align}

$z_i =  1-p_i$, $r_{ij}=0$, $x_i = \delta_i = 0$ for all $i,j$.

\underline{Region 4:}

\begin{align}
    \nonumber p_1 &= 1/2, \\
    \nonumber p_i &= \frac{1+(c_1+1)k(1-\beta)}{2} \quad \forall i>1.
\end{align}

$r_{1j} = c_1(1-p_2)-\frac{1}{2(n-1)} =\frac{c_1-c_1(c_1+1)k(1-\beta)}{2} - \frac{1}{2(n-1)}$ for all $j>1$ while $r_{ij} = 0$ for all $i>1$ and $j$.

$z_1 =(n-1)c_1(1-p_2) =\frac{1-(c_1+1)k(1-\beta)}{2}c_1(n-1)$, $z_i =  \frac{1-(c_1+1)k(1-\beta)}{2}$ for all $i>1$; $\delta_i = 0$ for all i.

Knowing all the optimal solutions of \eqref{eq:opt_c}, we can therefore compute the optimal profits by the objective function. We can then compare the profits of different regions to obtain the boundary values of $k$ that divide different regions. We denote the optimal profits of different regions as $\pi_{l}$ and the critical value of $k$ as $k_{l_1 \to l_2}$, where $l, l_1, l_2 \in {1,2,3,4}$ and $l_1 > l_2$. Hence $k_{l_1 \to l_2}$ represents the lowest value of $k$ that $\pi_{l_2} > \pi_{l_1}$ or optimal solution in region $l_1$ becomes infeasible. To justify Theorem \ref{thm:closed_form}, we need to find out the $k$ values each region will take part in.

\underline{$k_{4 \to 3}$: transition from $r>0$ to $r=0$ while $z>0$. }
\begin{equation}
     k_{4 \to 3} = \frac{(n-1)c_1-1}{(1-\beta)(n-1)(1+c_1)c_1}.
\end{equation}

There can be profit jump in this transition (since $k_{4 \to 3}$ is not obtained by profit equality $\pi_3 = \pi_4$ but instead, is the $k$ value when $r = 0$ from region 4).

\underline{$k_{3 \to 2}$, transition of $x=0$ to $x>0$ while $z>0$.}

\begin{equation}
    k_{3 \to 2} = \frac{(1+\beta)c_1 + (n-1)\beta c_1^3 + 1}{(c_1 + 1)(\beta(n-1) c_1^2 + 1)}.
\end{equation}

\underline{$k_{2 \to 1}$, transition of $z>0$ to $z=0$ while $x>0$.}
\begin{align*}
k_{2 \to 1} &= \begin{cases}
                  1 &\text{if }\xi \in [\frac{\beta(n-1)-1}{\beta(n-2)},1]\\
                  \frac{c_1(1+\beta) + (n-1)\beta^2 c_1^3 + 1}{(c_1 + 1)((n-1)\beta^2 c_1^2 + 1)} & \text{if } \xi \in [\beta_{lim} ,\frac{\beta(n-1)-1}{\beta(n-2)})\\
                  \frac{1+\beta c_1}{c_1 + 1} & \text{if }\xi \in [0,\beta_{lim}).
\end{cases}
\end{align*}

\underline{$k_{4 \to 2}$: transition from region 4 to region 2 directly.}
\begin{equation}
    k_{4 \to 2} = \frac{1+\beta c_1}{1+c_1}.
\end{equation}

\underline{$k_{4 \to 1}$: transition from region 4 to region 1 directly.}
\begin{equation}
    k_{4 \to 1} = \frac{1+\beta c_1}{1+c_1}.
\end{equation}

Since the $k_{4 \to 3}$, $k_{4 \to 2}$ and $k_{4 \to 1}$ are always real number in $[0,1]$, then there always exist some values of $k$ such that the optimal solution falls in region 4. Hence region 4 exists for any $n,\beta$ and $\xi$. Moreover, when $k$ grows infinitely large so that operating AV is much more expensive then HVs, then obviously the optimal solution will use HVs only for any possible $n, \beta$ and $\xi$. Therefore, region 1 also exist for all $n, \beta$ and $\xi$. However, as we will show later, there are values of $n,\beta$ and $\xi$ that the optimal solution of \eqref{eq:opt_d0} will never fall in region 2 or 3 for any $k$.

Region 3 exists means that region 4's solution transits to region 3 before reaching region 2 or 1. Also, $k_{4 \to 2} = k_{4 \to 1}$. Thus region 3 exists only if $k_{4 \to 3} < k_{4 \to 2}$. That is,
\begin{align}
    \nonumber   \frac{(n-1)c_1-1}{(1-\beta)(n-1)(1+c_1)c_1} &< \frac{1+\beta c_1}{1+c_1}\\
    \nonumber   (n-1)c_1-1 &< (1-\beta)(n-1)c_1(1+\beta c_1)\\
    \label{eq:num-1}   \beta c_1(n-1)(1-c_1+\beta c_1) &< 1.
\end{align}
Therefore, if the values of $n, \beta$ and $\xi$ do not satisfy \eqref{eq:num-1}, then region 3 does not exist and the optimal solution of \eqref{eq:opt_d0} will transit from region 4 directly to region 2 or 1. With this condition held, region 2 exists only if $k_{4 \to 2} < k_{2 \to 1}$ for similar reason as above.

If region 3 exists, then we claim that region 2 must exist because $k_{3 \to 2} \leq k_{2 \to 1}$ always.
To demonstrate that, we need to show that $\frac{(1+\beta)c_1 + (n-1)\beta c_1^3 + 1}{(c_1 + 1)(\beta(n-1) c_1^2 + 1)}\leq 1$, $\frac{(1+\beta)c_1 + (n-1)\beta c_1^3 + 1}{(c_1 + 1)(\beta(n-1) c_1^2 + 1)} \leq \frac{c_1(1+\beta) + (n-1)\beta^2 c_1^3 + 1}{(c_1 + 1)((n-1)\beta^2 c_1^2 + 1)}$ and that $\xi \notin [0,\beta_{lim})$ in this case. 

We will show the inequalities through contradiction.
Suppose first that
\begin{align}
    \nonumber \frac{(1+\beta)c_1 + (n-1)\beta c_1^3 + 1}{(c_1 + 1)(\beta(n-1) c_1^2 + 1)} &> 1\\
    \nonumber (1+\beta)c_1 + (n-1)\beta c_1^3 + 1 &> (c_1 + 1)(\beta(n-1) c_1^2 + 1)\\
    \nonumber \beta c_1 &> \beta(n-1)c_1^2\\
    \label{eq:num_2} 1 &> (n-1)c_1.
\end{align}
But $(n-1)c_1 = (n-1)+(2-n)\xi \in [0,1]$, hence $\frac{(1+\beta)c_1 + (n-1)\beta c_1^3 + 1}{(c_1 + 1)(\beta(n-1) c_1^2 + 1)}\leq 1$.

Similarly, suppose
\begin{align}
    \nonumber \frac{(1+\beta)c_1 + (n-1)\beta c_1^3 + 1}{(c_1 + 1)(\beta(n-1) c_1^2 + 1)} &> \frac{c_1(1+\beta) + (n-1)\beta^2 c_1^3 + 1}{(c_1 + 1)((n-1)\beta^2 c_1^2 + 1)} \\
    \nonumber \beta(\beta c_1+1) &> \beta c_1+1\\
    \nonumber \beta &>1.
\end{align}
Hence $\frac{(1+\beta)c_1 + (n-1)\beta c_1^3 + 1}{(c_1 + 1)(\beta(n-1) c_1^2 + 1)} \leq \frac{c_1(1+\beta) + (n-1)\beta^2 c_1^3 + 1}{(c_1 + 1)((n-1)\beta^2 c_1^2 + 1)}$.

When region 3 exists, then \eqref{eq:num-1} is satisfied. Solving \eqref{eq:num-1} gives that
\begin{align*}
 \xi > \max\bigg\{  &\frac{n-1}{2(1-\beta)\beta(n-2)}\bigg[\beta(1-2\beta)\\
 & + \sqrt{\frac{\beta^2(n-1)+\beta(4\beta-4)}{n-1}}\bigg],0\bigg\}\\
 > \beta_{lim}
\end{align*}
since $\beta \in (0,1)$. As a result, $\xi \notin [0,\beta_{lim})$ in this case.

As a result, $k_{3 \to 2} \leq k_{2 \to 1}$ when \eqref{eq:num-1} is satisfied and thus region 2 always exists if region 3 exists.

To summarize, if \eqref{eq:num-1} is satisfied, then all four regions exists, while the transition $k$ values are defined above. We then assume that \eqref{eq:num-1} is not satisfied, if $k_{4 \to 2} < k_{2 \to 1}$, then region 1, 2 and 4 exist while separated by $k_{2 \to 1}$ and $k_{4 \to 2}$; if  $k_{4 \to 2} \geq k_{2 \to 1}$, then only region 1 and 4 exist while separated by $k_{4 \to 1}$.

Define $k_1 = k_{4 \to 2}$, $k_2 = k_{2 \to 1}$, $k_3 = k_{4 \to 3}$ and $k_4 = k_{3 \to 2}$.

We can therefore conclude that for the case that  $k_3 \geq k_1$, suppose $k_1<k_2$, if $k \in [0,k_1]$, it is optimal for the platform to use only AVs; if $k \in (k_1,k_2)$, it is optimal to deploy a mixed-autonomy network; if $k \geq k_2$, it is optimal to use HVs only. Suppose $k_1 \geq k_2$, if $k \in [0, k_1]$, it is optimal to use AVs only and if $k> k_1$, it is optimal to use HVs only.

For the case that  $k_3 < k_1$, if $k \in [0,k_4]$, it is optimal for the platform to have only AVs; if $k \in (k_4,k_2)$, it is optimal to deploy a mixed-autonomy network; if $k \geq k_2$, it is optimal to use HVs only. 

\end{proof}
\end{proofcomment}

\begin{figure}[!t]
\centering
\includegraphics[width=.9\columnwidth]{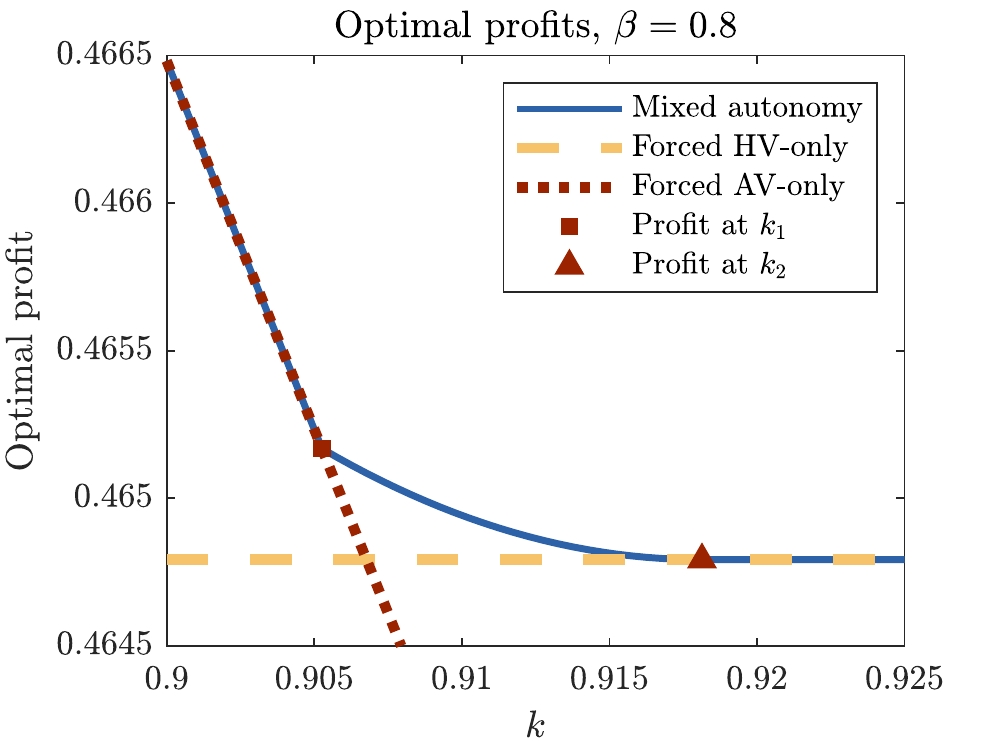}\\\includegraphics[width=.9\columnwidth]{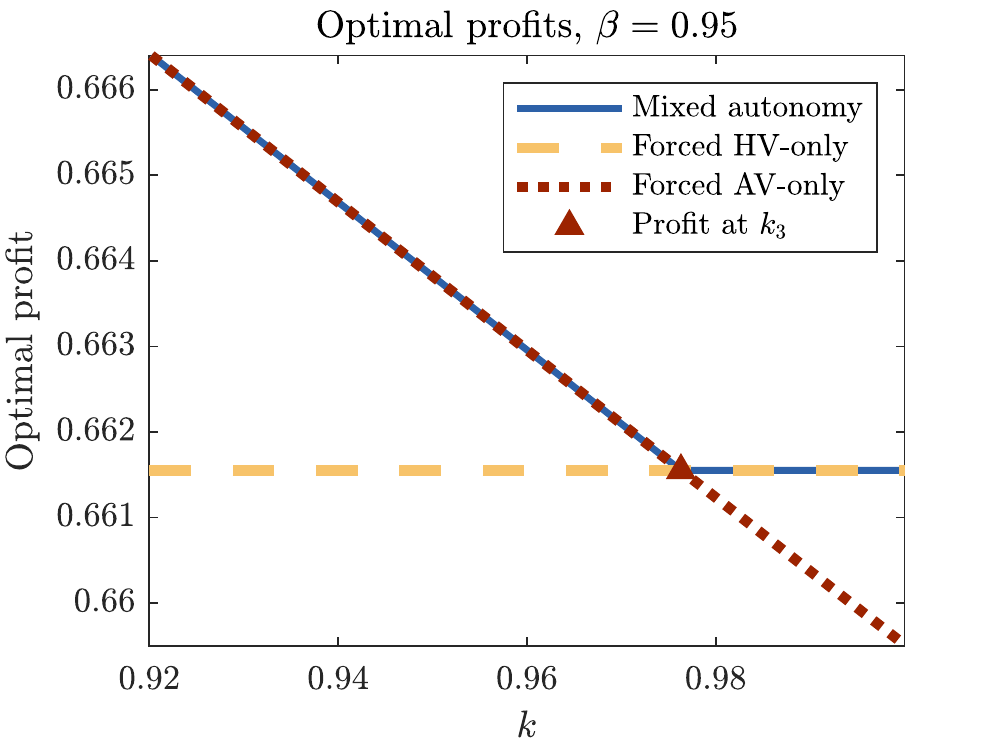}
\caption{Optimal profits for a star-to-complete network with $n=3$, $\xi=0.2$ under a mixed autonomy deployment, a forced HV-only deployment, and a forced AV-only deployment. (Top) When $\beta=0.8$, it is optimal for the platform to use only AVs when $k$, the ratio of the cost of AVs to HVs, satisfies $k\leq k_1=0.9053$, only HVs when $k\geq k_2=0.9181$, and a mix of AVs and HVs when $k_1<k<k_2$. (Bottom) When $\beta=0.95$, it is optimal for the platform to use only AVs when $k\leq k_1=0.9763$ and only HVs when $k > k_1$, and it is never optimal for the platform to use a mix of HVs and AVs.}
\label{fig1}
\end{figure}

\change{To demonstrate Theorem \ref{thm:closed_form} and provide intuition for the fundamental theoretical results of this paper, we study} a star-to-complete network with $n=3$, $\xi=0.2$. We consider two cases: $\beta=0.8$ and $\beta=0.95$, and we compute optimal equilibria and profits using the optimization problems formulated above. \change{In both cases, applying Theorem 5, we can verify that $k_3 \geq k_1$.} For the first case with $\beta=0.8$, we obtain $k_1=0.9053$ and $k_2=0.9181$ so that $k_1<k_2$. Figure \ref{fig1}(Top) confirms that for $k\leq k_1$, it is optimal for the platform to deploy only AVs, for $k_1<k<k_2$, it is optimal for the platform to use both AVs and HVs, and for $k\geq k_2$, it is optimal for the platform to use only HVs. In constrast, when $\beta=0.95$ so that the expected lifetime of HVs in the network is longer, then \change{$k_1=0.9763$ and} $k_1 \geq k_2$. Figure \ref{fig1} (Bottom) confirms that for \change{$k\leq k_1$}, the platform finds it optimal to deploy only AVs, and for \change{$k > k_1$}, the platform finds it optimal to use only HVs; there is no regime in which the platform finds it optimal to use both AVs and HVs. The plots in Figure \ref{fig1} are generated by solving the optimization problem \eqref{eq:opt_d2} in MATLAB using CVX, a package for specifying and solving convex programs\cite{cvx,gb08}. \change{Theorem \ref{thm:closed_form} guarantees that the basic qualitative results demonstrated here apply to arbitrarily large star-to-complete networks.} 

It is interesting to note from the above thresholds that even if AVs are cheaper than HVs, when the price difference is small, the platform may still choose to deploy only HVs or to deploy a mix of AVs and HVs. An explanation for this observation is as follows. Recall that with probability $1-\beta$, a driver leaves the network and does not seek to be matched to a new rider after finishing a ride and thus essentially provides one-way service. In contrast, AVs are assumed to remain in the network and must be recirculated to a new location. When the demand is uneven so that some destinations are more popular than others, the platform can exploit this one-way service to obtain a higher profit with HVs, even if AVs are less expensive on a per ride basis.

\section{Conclusion}
\label{sec:conclusions}
We proposed three models for ride-sharing systems with mixed autonomy under different ride-assigning schemes and showed that under equilibrium conditions, the optimal profits can be computed efficiently by converting the original problems into alternative convex programs. In addition, we proved that the optimal profits of the three models are the same.

We found that the optimal profits for the ride-sharing platform with AVs in the fleet will be the same as that of the human-only network when $k$ is large, \emph{i.e.}, the cost for operating an AV is relatively high compared to the outside option earnings for drivers' lifetime. In particular, in Proposition \ref{prop:lowerbound-general}, we showed that if the cost of operating an AV exceeds the expected compensation to a driver in the system, the platform will find it optimal to not use AVs, an intuitive result. 

The case study illustrates that the platform may not necessarily find it optimal to use AVs even when the cost of operating an AV is less than the expected compensation to a driver in the system. Moreover, there are some situations when it is optimal to have both drivers and AVs in the platform. We quantify the conditions for which the mixed autonomy deployment allows for higher profits than a forced AV-only or forced HV-only deployment.

The model proposed and studied here includes a several simplifying assumptions that can be relaxed in future work. For example, destinations are often not equidistant and ride costs might then depend on destination.  Nonetheless, these simplifying assumptions are important for illuminating fundamental properties of ride-sharing in a mixed autonomy setting.

\bibliography{Library}
\bibliographystyle{IEEEtran}

\end{document}